\documentclass[letterpaper,10pt,onecolumn,draftclsnofoot]{IEEEtran}

\usepackage{graphicx}
\usepackage{epsfig}
\usepackage{subfigure}
\usepackage{amssymb}
\usepackage{amsbsy}
\usepackage{amsmath}
\usepackage{cite}
\usepackage{url}

\usepackage{mathrsfs}

\usepackage{color}

\usepackage{placeins}
\usepackage{float}
\usepackage{tabularx,colortbl}
\usepackage{times,amsmath,epsfig}
\usepackage{xsp ace,latexsym,syntonly}
\usepackage{amssymb}
\usepackage{textcomp}
\usepackage{subfigure}
\usepackage{amsbsy}
\usepackage{cite}
\usepackage{verbatim}
\usepackage{bbm}
\usepackage{framed}
\usepackage{bbm,bm}
\usepackage{multirow}
\usepackage{mathtools}
\usepackage{algorithm} 
\usepackage{algorithmic} 
\usepackage{authblk}

\newcommand{\cA}{{\cal A}}

\newcommand{\cC}{{\cal C}}

\newcommand{\cU}{{\cal U}}
\newcommand{\cV}{{\cal V}}

\newtheorem{theorem}{Theorem}
\newtheorem{lemma}{Lemma}

\newtheorem{definition}{Definition}

\newcommand{\beq}{\begin{equation}}
\newcommand{\eeq}{\end{equation}}
\newcommand{\bea}{\begin{array}}
\newcommand{\ena}{\end{array}}
\newcommand{\bds}{\begin {itemize}}
\newcommand{\eds}{\end {itemize}}
\newcommand{\bdf}{\begin{definition}}
\newcommand{\blm}{\begin{lemma}}
\newcommand{\edf}{\end{definition}}
\newcommand{\elm}{\end{lemma}}
\newcommand{\bthm}{\begin{theorem}}
\newcommand{\ethm}{\end{theorem}}
\newcommand{\bprp}{\begin{prop}}
\newcommand{\eprp}{\end{prop}}
\newcommand{\bcl}{\begin{claim}}
\newcommand{\ecl}{\end{claim}}
\newcommand{\bcr}{\begin{coro}}
\newcommand{\ecr}{\end{coro}}
\newcommand{\bquest}{\begin{question}}
\newcommand{\equest}{\end{question}}

\newcommand{\larrow}{{\larrow}}

\def\urltilda{\kern -.15em\lower .7ex\hbox{\~{}}\kern .04em}

\begin{document}\title{Information-Directed Random Walk for Rare Event Detection in Hierarchical Processes}

%

\author{Chao Wang, Kobi Cohen, Qing Zhao
\thanks{Chao Wang and Qing Zhao are with the School of Electrical and Computer Engineering, Cornell University. Emails: \{cw733, qz16\}@cornell.edu. Kobi Cohen is with the Department of Electrical and Computer Engineering, Ben-Gurion University of the Negev. Email: yakovsec@bgu.ac.il.}
\thanks{The work of C. Wang and Q. Zhao was supported by the U.S. Army Research Office under Grant W911NF-17-1-0464 and the National Science Foundation under Grant CCF-1815559. The work of K. Cohen was supported by the Cyber Security Research Center at Ben-Gurion University of the Negev under grant 07616 and the U.S.-Israel Binational Science Foundation (BSF) under grant 2017723. Part of the results have been presented at the 2017 IEEE International Symposium on Information Theory and the 2018 IEEE International Workshop on Signal Processing Advances in Wireless Communications.}
}
\maketitle

%
\begin{abstract}
\label{sec:abstract}
The problem of detecting a few anomalous processes among a large number of data streams is considered. At each time, aggregated observations can be taken from a chosen subset of the processes, where the chosen subset conforms to a given tree structure. The random observations are drawn from a general distribution that may depend on the size of the chosen subset and the number of anomalous processes in the subset. We propose a sequential search strategy by devising an information-directed random walk (IRW) on the tree-structured observation hierarchy. Subject to a reliability constraint, the proposed policy is shown to be asymptotically optimal with respect to the detection accuracy. Furthermore, it achieves the optimal logarithmic-order sample complexity with respect to the size of the search space provided that the Kullback-Liebler divergence between aggregated observations in the presence and the absence of anomalous processes are bounded away from zero at all levels of the tree structure as the size of the search space approaches infinity. Sufficient conditions on the decaying rate of the aggregated observations to pure noise under which a sublinear scaling in the size of the search space is preserved are also identified for the Bernoulli case.
\end{abstract}
%
\def\keywords{\vspace{.5em}
{\bfseries\textit{Index Terms}---\,\relax%
}}
\def\endkeywords{\par}
\keywords
Sequential design of experiments, active hypothesis testing, anomaly detection, noisy group testing, channel coding with feedback.
\section{Introduction}
\label{sec:intro}

\subsection{Rare Event Detection in Hierarchical Processes} 
\label{ssec:search_rare}

Consider the problem of detecting anomalies in a large number of processes. At each time, the decision maker chooses a subset of processes to observe. The chosen subset conforms to a predetermined tree structure. The (aggregated) observations are drawn from a general distribution that may depend on the size of the chosen subset and the number of anomalies in the subset. The objective is a sequential search strategy that adaptively determines which node on the tree to probe at each time and when to terminate the search in order to minimize a Bayes risk that takes into account both the sample complexity and the detection accuracy.

The above problem is an archetype for searching for a few rare events of interest among a massive number of data streams that can be observed at different levels of granularity. For example, financial transactions can be aggregated at different temporal and geographic scales. In computer vision applications such as bridge inspection by UAVs with limited battery capacity, sequentially determining areas to zoom in or zoom out can quickly locate anomalies by avoiding giving each pixel equal attention. A particularly relevant application is heavy hitter detection in Internet traffic monitoring. It is a common observation that Internet traffic flows are either ``elephants'' (heavy hitters) or ``mice'' (normal flows). A small percentage of high-volume flows account for most of the total traffic~\cite{Thompson_wide}. Quickly identifying heavy hitters is thus crucial to network stability and security, especially in detecting denial-of-service (DoS) attacks. Since maintaining a packet count for each individual flow is infeasible due to limited sampling resources at the routers, an effective approach is to aggregate flows based on the IP prefix of the source or destination addresses\cite{cormode2003finding, QGT_TSP18}. Indeed, recent advances in software-defined networking (SDN) allow programmable routers to count aggregated flows that match a given IP prefix. The search space of all traffic flows thus follows a binary tree structure.

\subsection{Information-Directed Random Walk}

To fully exploit the hierarchical structure of the search space, the key questions are how many samples to obtain at each level of the tree and when to zoom in or zoom out on the hierarchy. A question of particular interest is whether a sublinear scaling of the sample complexity with the size of the search space is feasible while achieving the optimal scaling with the detection accuracy. In other words, whether accurate detection can be achieved by examining only a diminishing fraction of the search space as the search space grows.

Our approach is to devise an information-directed random walk (IRW) on the hierarchy of the search space. The IRW initiates at the root of the tree and eventually arrives and terminates at the targets (i.e., the anomalous processes) with the required reliability. Each move of the random walk is guided by the test statistic of the sum log-likelihood ratio (SLLR) collected from each child of the node currently being visited by the random walk. This local test module ensures that the global random walk is more likely to move toward a target than move away from it and that the walk terminates at a true target with the required detection accuracy. \par

Analyzing the sample complexity of the IRW strategy lies in examining the trajectory of the biased random walk. With a suitably chosen confidence level in the local test module, the random walk will concentrate, with high probability, on a smaller and smaller portion of the tree containing the targets and eventually probes the targets only. The basic structure of the analysis is to partition the tree into a sequence of half trees with decreasing size, and bound the time the random walk spent in each half tree. The entire search process, or equivalently, each sample path of the biased random walk, is then partitioned into stages by the successively defined last passage time to each of the half trees in the shrinking sequence. We show that the sample complexity of the IRW strategy is asymptotically optimal in detection accuracy and order optimal, specifically, a logarithmic order, in the size of the search space when the aggregrated observations are informative at all levels of the tree.

We also consider the case when higher-level observations decay to pure noise. Using Bernoulli distribution as a case study, we show that when the Kullback-Leibler (KL) divergence between the target-absent and target-present distributions decays to 0 in polynomial order with the depth of the tree, the IRW offers a sample complexity that is poly-logarithmic order in the number of processes; when the decay rate is exponential in the level $l$ of the tree (i.e., $\propto \alpha^{-2l}$), a sublinear scaling in the size of the search space
can be achieved provided that $1<\alpha<\sqrt{2}$.

The proposed search strategy is deterministic with search actions explicitly specified at each given time. It involves little online computation beyond calculating the SLLR and performing simple comparisons. The proposed strategy is also efficient in terms of computation and memory requirement. By effectively localizing the data processing to small subsets of the search space, it has $O(1)$ computation and memory complexity.

\subsection{Related Work}
\label{ssec:related_work}

The problem considered here falls into the general class of sequential design of experiments pioneered by Chernoff in 1959~\cite{chernoff1959sequential} in which he posed a binary active hypothesis testing problem. Compared with the classic sequential hypothesis testing pioneered by Wald \cite{wald1947} where the observation model under each hypothesis is fixed, active hypothesis testing has a control aspect that allows the decision maker to choose different experiments (associated with different observation models) at each time. Chernoff proposed a \emph{randomized} strategy and showed that it is asymptotically optimal as the error probability approaches zero. Known as the Chernoff test, this randomized strategy chooses, at each time, a probability distribution governing the selection of experiments based on all past actions and observations. The probability distribution is given as a solution to a maxmin problem that can be difficult to solve, especially when the number of hypotheses and/or the number of experiments is large, a case of focus in this paper with a large number of processes. Furthermore, the Chernoff test does not address the scaling with the number of hypotheses and results in a linear sample complexity in the size of the search space when applied to the problem considered here.\par

A number of variations and extensions of Chernoff's randomized test have been considered (see, for example,~\cite{bessler1960theory, nitinawarat2013controlled, naghshvar2013active}). In particular, in~\cite{naghshvar2013active},  Naghshvar and Javidi developed a randomized test that achieves the optimal logarithmic order of the sample complexity in the number of hypotheses under certain implicit conditions. These conditions, however, do not hold for the problem considered here. Furthermore, similar to the Chernoff test, this randomized test is specified only implicitly as solutions to a sequence of maxmin problems that can be intractable for general observation distributions and large problem size.

The problem of detecting anomalies or outlying sequences has also been studied under different formulations, assumptions, and objectives~\cite{heydari2016quickest,tajer2013quick, geng2013quickest}. An excellent survey can be found in~\cite{Tajer&etal:2014survey}. These studies, in general, do not address the optimal scaling in both the detection accuracy and the size of the search space.

The problem studied here also has intrinsic connections with several other problems studied in different application domains, in particular, adaptive sampling, noisy group testing and compressed sensing, and channel coding with feedback. We discuss the connections and differences in Section~\ref{sec:connections}.

\section{Problem Formulation}
\label{sec:formulation}

We first focus on the problem of detecting a single target. The problem of detecting multiple targets is discussed in Section~\ref{sec:multi_target}.

\begin{figure}[t!]
      \centering
      \includegraphics[scale=0.25]{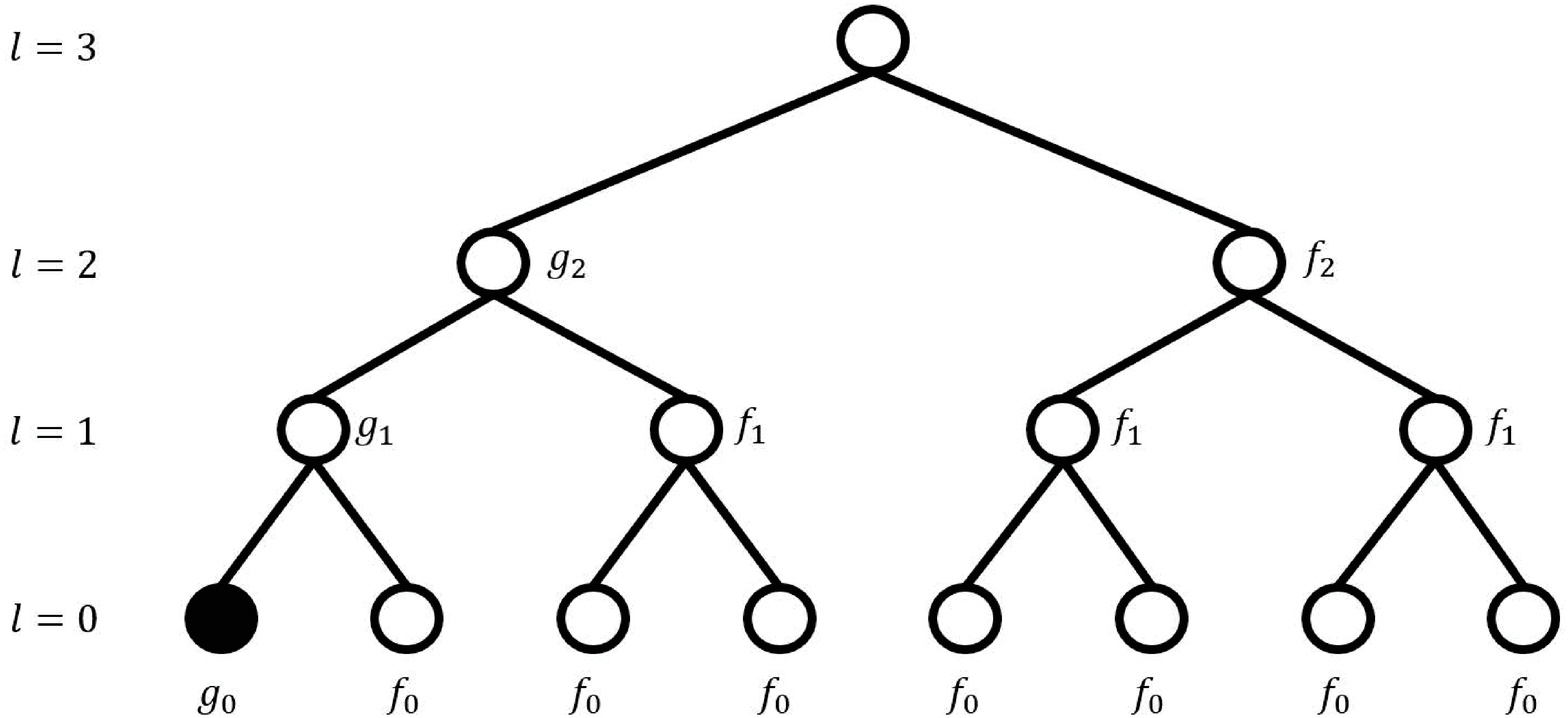}
      \caption{A binary tree observation model with a single target.}
      \label{fig:tree_single}
   \end{figure}

For the ease of presentation, we focus on the case of a binary tree structure as illustrated in Fig.~\ref{fig:tree_single}. Extensions to a general tree structure are straightforward. Let $g_0$ and $f_0$ denote, respectively, the distributions of the anomalous process and the normal processes\footnote{The proposed policy and the analysis extend with simple modifications to cases where each process has different target-present and target-absent distributions.}. Let $g_l$~($l=1,\ldots,\log_2 M$) denote the distribution of the measurements that aggregate the anomalous process and $2^l-1$~normal processes, and $f_l$~($l=1,\ldots,\log_2 M$) denote the distribution of the measurements that aggregate $2^l$ normal processes~(see~Fig.~\ref{fig:tree_single}). We allow general relation between $\{g_l,\,f_l\}$ and $\{g_0,\, f_0\}$, which often depends on the specific application. For example, in the case of heavy hitter detection where the measurements are packet counts of an aggregated flow, $g_l$ and $f_l$ are given by multi-fold convolutions of $f_0$ and $g_0$. For independent Poisson flows, $g_l$ and $f_l$ are also Poisson with mean values given by the sum of the mean values of their children at the leaf level. As is the case in most of the practical applications, we expect that observations from each individual process are more informative than aggregated observations. More precisely, we expect $D(g_{l-1}\|f_{l-1})\geq D(g_l\|f_l)$ and $D(f_{l-1}||g_{l-1})\geq D(f_l\|g_l)$ for all $l>0$, where $D(\cdot||\cdot)$ denotes the KL divergence between two distributions. However, the results in this work hold for the general case without the monotonicity assumptions above. \par

We aim to develop an active search strategy that sequentially determines whether to terminate the search and if not, which node on the tree to probe next.  Specifically, an active search strategy $\Gamma=(\{\phi(t)\}_{t\ge 1},\, \tau, \, \delta)$ consists of a sequence of selection rules $\{\phi(t)\}_{t\ge 1}$ governing which node to probe at each time, a stopping rule $\tau$ deciding when to terminate the search, and a declaration rule $\delta$ deciding which leaf node is the target at the time of stopping.

We adopt the Bayesian approach as in Chernoff's original work~\cite{chernoff1959sequential} and subsequent studies in~\cite{naghshvar2013active, Cohen2015active}. Specifically, a sampling cost of $c\in(0,1)$ is incurred for each observation and a loss of $1$ a wrong declaration. Let $\pi_m$ denote the \emph{a priori} probability that process~$m$ is anomalous, which is referred to as hypothesis $H_m$. The probability of detection error $P_e(\Gamma)$ and the sample complexity $Q(\Gamma)$ of strategy $\Gamma$ are given by
\begin{eqnarray}
P_{e}(\Gamma) &\triangleq& \sum_{m=1}^M \pi_m \mathbb{P}{_\Gamma} \left[\delta \neq m | H_m \right],\\
Q(\Gamma) &\triangleq& \sum_{m=1}^M \pi_m \mathbb{E}_\Gamma\left[\tau | H_m\right],
\end{eqnarray}
where $\mathbb{P}_\Gamma$ and $\mathbb{E}_\Gamma$ denote the probability measure and expectation with respect to the probability space induced by $\Gamma$. The dependency on $\Gamma$ will be omitted from the notations when there is no ambiguity. The Bayes risk of $\Gamma$ is then given by
\begin{equation}\label{risk}
R(\Gamma) \triangleq P_{e}(\Gamma) + cQ(\Gamma).
\end{equation} The objective is a strategy $\Gamma$ that achieves the lower bound of the Bayes~risk:
\begin{equation}
R^* = \displaystyle\inf_{\Gamma} \, R(\Gamma).
\end{equation}

We are interested in strategies that offer the optimal scaling in both $c$, which controls the detection accuracy, and $M$, which is the size of the search space. 
A test $\Gamma$ is said to be \emph{asymptotically optimal} in $c$ if, for fixed $M$,
\begin{equation}\label{asy_opt}
\lim_{c\rightarrow 0}\ \frac{R(\Gamma)}{R^*} = 1.
\end{equation}
A shorthand notation $R(\Gamma)\sim R^*$ will be used to denote the relation specified in~\eqref{asy_opt}. If the above limit is a constant greater than $1$, then $\Gamma$ is said to be order optimal.
The asymptotic and order optimalities in $M$ are similarly defined as $M$ approaching infinity for a fixed~$c$. 

A dual formulation of the problem is to minimize the sample complexity subject to an error constraint $\varepsilon$, i.e.,
\begin{equation}\label{error_formula}
\Gamma^* = \arg\displaystyle\inf_{\Gamma} \, Q(\Gamma), ~~~s.t.~~~ P_e(\Gamma)\le \varepsilon.
\end{equation}

In the Bayes risk given in \eqref{risk}, $c$ can be viewed as the inverse of the Lagrange multiplier, thus controls the detection accuracy of the test that achieves the minimum Bayes risk. Following the same lines of argument in~\cite{Lai:1988nearly, Shiryaev:2007optimal}, one can obtain the solution to~\eqref{error_formula} from the solution under the Bayesian formulation.

\section{Information-Directed Random Walk}
\label{sec:IRW_policy}

The IRW policy induces a biased random walk that initiates at the root of the tree and eventually arrives at the target with a sufficiently high probability. Each move of the random walk is guided by the output of a local test carried on the node currently being visited by the random walk. This local test module $\mathcal{L}(p)$ determines, with a confidence level $p$, whether this node contains the target, and if yes, which child contains the target. Based on the output, the random walk zooms out to the parent\footnote{The parent of the root node is defined as itself.} of this node or zooms in to one of its child. The confidence level $p$ of the local test module is set to be greater than $\frac{1}{2}$ to ensure that the global random walk is more likely to move toward the target than move away from it. \par

Once the random walk reaches a leaf node, say node $m$ ($m = 1,\ldots,M$), samples are taken one by one from node $m$ and the SLLR $S_m(t)$ of node $m$ is
updated with each new sample taken during the current visit to this node:
\begin{equation}\label{inSLLR}
S_m(t) = \sum_{n=1}^{t} \log \frac{g_{0}(y(n))}{f_{0}(y(n))}.
\end{equation}
When $S_m(t)$ drops below $0$, the random walk moves back to the parent of node $m$. When $S_m(t)$ exceeds $\log\frac{\log_2 M}{c}$, the detection process terminates, and node $m$ is declared as the target. The choice of the stopping threshold $\log\frac{\log_2 M}{c}$ is to ensure that the error probability is in the order of $O(c)$, which in turn secures the asymptotic optimality in $c$ (see Theorem~\ref{main_thm} and the proof in Appendix~\ref{appx:proof_thm1}).

We now specify the local test module $\mathcal{L}(p)$ carried out on upper-level nodes. The objective of $\mathcal{L}(p)$ is to distinguish three hypotheses -- $H_0$ that this node does not contain the target and $H_1$ ($H_2$) that the left (right) child of this node contains the target -- with a confidence level no smaller than $p$ under each hypothesis. Various tests (fixed-sample-test, sequential, and active) with a guaranteed confidence level $p$ can be constructed for this ternary hypothesis testing problem. We present below a fixed-sample-size test. A sequential test based on the Sequential Probability Ratio Test (SPRT)~\cite{wald1945} and an active test can also be constructed. Details are given in Appendix~\ref{appx:seq_local}. While the fixed-sample-size test specified below suffices to safeguard the asymptotic/order optimalities of IRW as given in Theorem~\ref{main_thm}, using the sequential and active local test improves the performance in the finite regimes as demonstrated in simulation examples in Section~\ref{sec:simulation}.

   \begin{figure}[b!]
      \centering
      \includegraphics[scale=0.25]{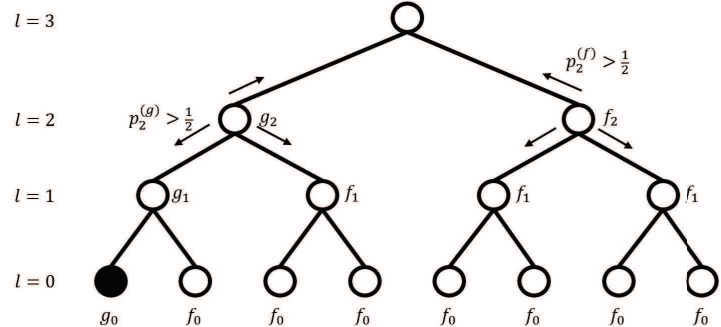}
      \caption{A biased random walk on the tree.}
      \label{fig:biased_RW}
   \end{figure}

Suppose that the random walk is currently at a node on level $l>0$.  A fixed $K_l$ samples, denoted as $y(n)$ ($n=1,\ldots, K_l$), are taken from each child of the node. The SLLR of each child is computed:
\begin{equation}\label{single_SLLR}
\sum_{n=1}^{K_l}\log \frac{g_{l-1}(y(n))}{f_{l-1}(y(n))}.
\end{equation}
If the SLLRs of both children are negative, the local test declares hypothesis $H_0$. Otherwise, the local test declares $H_1$ ($H_2$) if the left (right) child has a larger SLLR. The sample size $K_l$ is chosen to ensure a probability $p>\frac{1}{2}$ of correct detection under each of the three hypotheses and can be determined as follows. Let $p_l^{(g)}$ and $p_l^{(f)}$ denote, respectively, the probability that the local test output moves the random walk closer to the target when this node contains the target and when it does not~(see Fig.~\ref{fig:biased_RW}). Both are functions of the sample size $K_l$. We have
\begin{equation}\label{pg_pf}
\begin{split}
p_l^{(g)}&=\Pr\Bigg[ \sum_{n=1}^{K_l}  \log \frac{g_{l-1}(Y_n)}{f_{l-1}(Y_n)}>\max\Big\lbrace \sum_{n=1}^{K_l}  \log \frac{g_{l-1}(Z_n)}{f_{l-1}(Z_n)}, 0\Big\rbrace \Bigg], \\
p_l^{(f)}&=\left[\Pr\left(\sum_{n=1}^{K_l}  \log \frac{g_{l-1}(Z_n)}{f_{l-1}(Z_n)}<0\right)\right]^2,
\end{split}
\end{equation}
where $\lbrace Y_n\rbrace_{n=1}^{K_l}$ and $\lbrace Z_n\rbrace_{n=1}^{K_l}$ are i.i.d. random variables with distribution $g_{l-1}$ and $f_{l-1}$, respectively. The parameter $K_l$ ($l=1,2,\ldots, \log_2 M$) is chosen to ensure that $p_l^{(g)}>p$ and $p_l^{(f)}>p$. Note that the value of $K_l$ can be computed offline and simple upper bounds suffice.

\section{Performance Analysis}
\label{sec:performance}

In this section, we establish the asymptotic optimality of the IRW policy in $c$ and the order optimality in $M$.

\subsection{Main Structure of the Analysis}
\label{ssec:main_structure}

\begin{figure}[t!]
      \centering
      \includegraphics[scale=0.4]{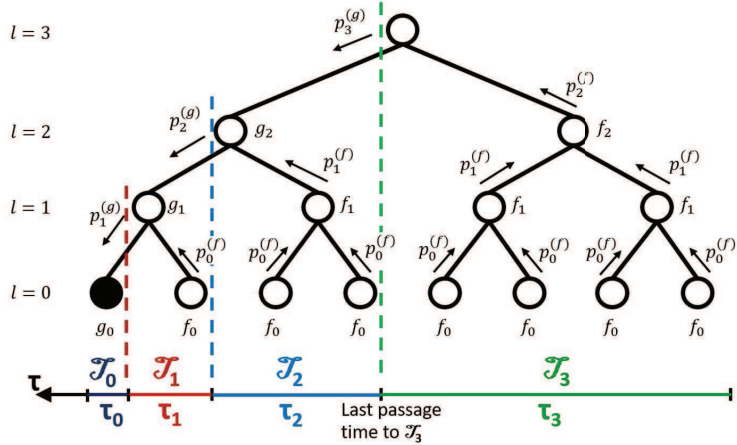}
      \caption{Partition of the tree into $\log_2 M +1$ half trees.}
      \label{fig:exit_time_single}
   \end{figure}

Analyzing the Bayes risk of the IRW strategy lies in
examining the trajectory of the biased random walk. With a confidence level $p>\frac{1}{2}$ in the
local test module, the random walk will concentrate, with high probability, on a smaller and smaller
portion of the tree containing the target and eventually probes the target only. Based on this insight, out approach is to partition the tree into $\log_2 M +1$ half trees $\mathscr{T}_{\log_2M}$, $\mathscr{T}_{\log_2M-1}$, $\ldots$, $\mathscr{T}_0$ with decreasing size, and bound the time the random walk spent in each half tree. As illustrated in Fig.~\ref{fig:exit_time_single} for $M=8$, $\mathscr{T}_l$ is the half tree (including the root) rooted at level $l$  ($l=\log_2 M,\ \log_2 M-1,\ldots,\ 1$) that does not contain the target and $\mathscr{T}_0$ consists of only the target node. The entire search process, or equivalently, each sample path of the resulting random walk, is then partitioned into $\log_2 M +1$ stages by the successively defined \emph{last passage time} to each of the half trees in the shrinking sequence. In particular, the first stage with length $\tau_{\log_2 M}$ starts at the beginning of the search process and ends at the last passage time to the first half tree $\mathscr{T}_{\log_2 M}$ in the sequence, the second stage with length $\tau_{\log_2 M -1}$ starts at $\tau_{\log_2 M}+1$ and ends at the last passage time to $\mathscr{T}_{\log_2 M -1}$, and so on. Note that if the random walk terminates at a half tree $\mathscr{T}_l$ with $l>0$ (i.e., a detection error occurs), then $\tau_j=0$ for $j=l-1,\ldots, 0$ by definition. It is easy to see that, for each sample path, we have the total time of the random walk equal to $\sum_{l=0}^{\log_2 M} \tau_l$.

\subsection{Informative Observations at All Levels}
\label{ssec:informative}
We now give the detailed analysis by first considering the scenario where the KL divergence between aggregated
observations in the presence and the absence of anomalous
processes is bounded away from zero at all levels of the tree. The theorem below characterizes the Bayes risk of the IRW policy. 
\begin{theorem}\label{main_thm}
Assume that there exists a constant $\delta>0$ independent of $M$ such that $D(g_l\|f_l) > \delta$ and $D(f_l\|g_l) > \delta$ for all $l=1,2,\ldots, \log_2 M$. For all $M$ and $c$, we have
\begin{eqnarray}\label{IRW_risk}
R(\Gamma_{\mbox{\footnotesize IRW}})\leq cB\log_2 M +\frac{c\log \frac{\log_2 M}{c}}{D(g_0\|f_0)} + O(c),
\end{eqnarray}
where $B$ is a constant independent of $c$ and $M$. Furthermore, the Bayes risk of IRW is order optimal in~$M$ for all $c$ and asymptotically optimal in $c$ for all $M$ greater than a finite constant $M_0$.
\end{theorem}
\begin{IEEEproof}
See Appendix~\ref{appx:proof_thm1}. 
\end{IEEEproof}

\medskip
The optimality of the Bayes risk of IRW in both $c$ and $M$ directly carries through to the sample complexity of IRW. Specifically, from \eqref{IRW_risk}, we have the following upper bound on the sample complexity of the IRW policy
\begin{equation}\label{IRW_et}
Q\left({\Gamma_{\mbox{\footnotesize IRW}}}\right)  \leq B\log_2 M +\frac{\log \frac{\log_2 M}{c}}{D(g_0\|f_0)} + O(1).
\end{equation}
For a fixed $M$, we readily have 
$$Q\left({\Gamma_{\mbox{\footnotesize IRW}}}\right) \sim \frac{-\log c}{D(g_0\|f_0)}.$$
For a fixed $c>0$, we have
$$Q\left({\Gamma_{\mbox{\footnotesize IRW}}}\right) = O(\log_2 M).$$
Comparing with the lower bound developed in~\cite{Cohen2015active} and~\cite{naghshvar2013active}, the sample complexity of IRW is asymptotically optimal in $c$ and order optimal in $M$.

\subsection{Aggregated Observations Decay to Pure Noise}
\label{ssec:pure_noise}

When the quality of higher level measurements decays sufficiently fast, the sample size $K_l$ of the local test may increase unboundedly with $l$ ($l=1,\ldots,$). Nevertheless, since the number of levels is $\log_2 M$, a sublinear scaling with $M$ is still attainable at moderate decaying rate of the aggregated observations. Using Bernoulli distribution as a case study, we examine the scenario where higher level observations decay to pure noise as $M$ grows. We establish sufficient conditions on the decaying rate of the quality of the hierarchical observations under which the IRW policy achieves a sublinear sample complexity in~$M$.

Assume that $f_l$ and $g_l$ follow Bernoulli distributions with parameters $u_l$ and $1-u_l$, respectively. In other words, the false alarm and miss detection probabilities at level $l$ are given by $u_l$. The KL divergence between $g_l$ and $f_l$ is $D(g_l\|f_l)=D(f_l\|g_l) = (1-2\mu_l) \log \frac{1-\mu_l}{\mu_l}$. We consider the case that $\mu_l$ increases with $l$ and converges to $\frac{1}{2}$ as $M$ approaches infinity. In this case, both $D(g_l\|f_l)$ and $D(f_l\|g_l)$ converge to zero, which leads to unbounded $K_l$. The following two theorems characterize the sample complexity of IRW when $\mu_l$ converges to $\frac{1}{2}$ in polynomial order and exponential order, respectively.

\medskip

\begin{theorem}\label{poly}
Assume that $\mu_l = \frac{1}{2} - (\frac{1}{2}-\mu_0)(l+1)^{-\alpha}$ ($l=0,1,2,\ldots, \log_2 M$) for some $\alpha\in\mathbb{Z}^+$ and $\mu_0<\frac{1}{2}$. The Bayes risk of the IRW policy is upper bounded by:
\begin{equation}\label{delay_poly}
R\left({\Gamma_{\mbox{\footnotesize IRW}}}\right) \leq O(c(\log_2 M)^{2\alpha+1}) +\frac{c\log \frac{\log_2 M}{c}}{D(g_0\|f_0)} + O(c). 
\end{equation}
\end{theorem}
\medskip
\begin{IEEEproof}
See Appendix~\ref{appx:proof_thm23}.  
\end{IEEEproof}

The case specified in Theorem~\ref{poly} corresponds to a polynomial decay of the KL divergence: $D(g_l\|f_l)=D(f_l\|g_l)\propto (l+1)^{-2\alpha}$. In this case, IRW offers a sample complexity that is poly-logarithmic in $M$:
$$Q\left({\Gamma_{\mbox{\footnotesize IRW}}}\right)= O((\log_2 M)^{2\alpha+1}).$$

\begin{theorem}\label{exp}
Assume that $\mu_l = \frac{1}{2} - (\frac{1}{2}-\mu_0)\alpha^{-l}$ ($l=0,1,2,\ldots, \log_2 M$) for some $\alpha >1$ and $\mu_0<\frac{1}{2}$.   
The Bayes risk of the IRW policy is upper bounded by:
\begin{equation}\label{delay_exp}
R\left({\Gamma_{\mbox{\footnotesize IRW}}}\right) \leq c\tilde{B}M^{{\log_2 \alpha^2}} +\frac{c\log \frac{\log_2 M}{c}}{D(g_0\|f_0)} + O(c), 
\end{equation}
where $\tilde{B}$ is a constant independent of $c$ and $M$. 
\end{theorem}

\begin{IEEEproof}
See Appendix~\ref{appx:proof_thm23}. 
\end{IEEEproof}

\medskip

The case specified in Theorem~\ref{exp} corresponds to a exponential decay of the KL divergence: $D(g_l\|f_l)=D(f_l\|g_l)\propto  \alpha^{-2l}$. In this case, IRW offers a sample complexity that is sub-liner in $M$ provided that $1<\alpha<\sqrt{2}$:
$$Q\left({\Gamma_{\mbox{\footnotesize IRW}}}\right)= O(M^{{\log_2 \alpha^2}}).$$

\section{Multi-Target Detection}
\label{sec:multi_target}

We now consider the problem of detecting~$L$ ($L\ge 0$) anomalous processes. We show that an extension of the IRW policy preserves the asymptotic optimality in $c$ and the order optimality in $M$ even when the number $L$ of targets is unknown.

\begin{figure}[t!]
      \centering
      \includegraphics[scale=0.25]{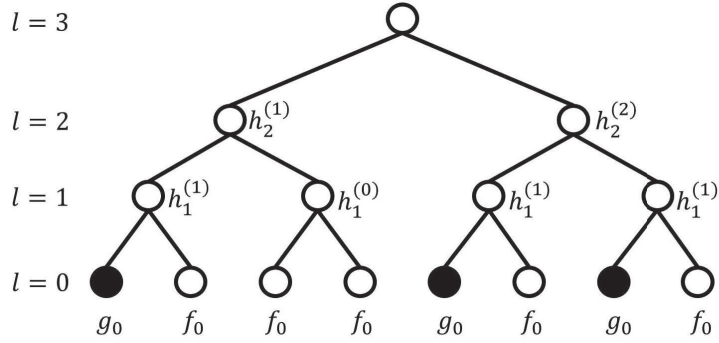}
      \caption{A binary tree observation model with multiple targets.}
      \label{fig:tree_multi}
   \end{figure}

Let $h_l^{(d)}$ ($l=0,1,\ldots, \log_2 M$, $d\leq \min\{L, 2^{l}\}$) denote the distribution of the measurements that aggregate $d$ anomalous processes and $2^l-d$ normal processes. An example with $M=8$ and $L=3$ is shown in Fig.~\ref{fig:tree_multi}. For a given $d$, we assume that for any $d'\leq d-1$, we have
\begin{equation}\label{KL_assump1}
D\left(h_{l-1}^{(d)}\big\| h_{l-1}^{(d')}\right) - D\left(h_{l-1}^{(d)}\big\| h_{l-1}^{(d'+1)}\right) >0,
\end{equation}
and for any $d'\geq d$, we have 
\begin{equation}\label{KL_assump2}
D\left(h_{l-1}^{(d)}\big\| h_{l-1}^{(d')}\right) - D\left(h_{l-1}^{(d)}\big\| h_{l-1}^{(d'+1)}\right)<0.
\end{equation}
The above monotonicity assumption on the KL divergence between $h_{l}^{(d)}$ and $h_{l}^{(d')}$ implies that a bigger difference $|d-d'|$ in the number of targets contained in the aggregated measurements leads to more distinguishable observations, which is usually true in practice. For the analysis in this section, we focus on the case that observations at all levels are informative. In other words, we assume that there exists a constant $\delta>0$ independent of $M$ such that $D(h_l^{(d+k)}\|h_l^{(d)})>\delta$ for all $l=1,2,\ldots,\log_2 M$, $d=0,1,\ldots \min \lbrace L, 2^l \rbrace$, and $-d\leq k\leq\min\lbrace L, 2^l\rbrace -d$. 

\subsection{IRW Policy for Known $L$}

We first consider the case that the number of targets is known. The IRW policy locates the $L$ ($L>1$) targets one by one\footnote{We do not assume that declared targets can be removed thus excluded from future observations.}. Similar to the single-target case, a biased random walk initiates at the root of the tree and eventually arrives at an undeclared target with high probability (referred to as one run of the random walk). The random walk is then reset to the root until $L$ targets have been declared.

The local test $\mathcal{L}(p)$ on an upper-level node differs from the single-target case in that it now faces four hypotheses, with an addition of hypothesis $H_3$ that both children contain undeclared targets. The outputs of $H_0$, $H_1$, and $H_2$ of the local test guide the random walk in the same way as in the single-target case. When the local test outputs $H_3$, the random walk arbitrarily chooses one child to zoom in. The stopping rule and declaration rule at each run of the random walk remain the same as in the single-target case.


To specify the local test module $\mathcal{L}(p)$, assume that the random walk is currently at a node on an upper level $l>0$, whose left and right child contain, respectively, $\widehat{d}_L$ and $\widehat{d}_R$ declared targets. Note that the local test faces a composite hypothesis testing problem, since both the left and right children may contain more than one target. Consider a fixed-sample test where $K_l^{(\widehat{d}_L)}$ and $K_l^{(\widehat{d}_R)}$ samples are taken from the left and right child, respectively. Note that the number of samples taken from each child depends on the number of declared targets. The SLLR of left child is computed as
\begin{equation}\label{SLLR_multi}
\sum_{n=1}^{K_l^{(\widehat{d}_L)}} \log \frac{h_{l-1}^{(\widehat{d}_L+1)}(y(n))}{h_{l-1}^{(\widehat{d}_L)}(y(n))}.
\end{equation}
The SLLR of the right child is similarly obtained. Based on the monotonicity assumption specified in~\eqref{KL_assump1} and~\eqref{KL_assump2}, the expected value of the log-likelihood ratio (LLR) of each sample in~\eqref{SLLR_multi} is positive when there are undeclared targets contained in the tested child, and is negative otherwise. The local decision rule is thus based on whether the SLLRs of the two children are both negative ($H_0$), both positive ($H_3$), or one negative one positive ($H_1$ or $H_2$). Similar to the single-target case, the values of $K_l^{(\widehat{d}_L)}$ and $K_l^{(\widehat{d}_R)}$ are chosen to guarantee the probability of declaring the correct hypothesis is greater than $p$. A sequential local test and an active local test can be similarly obtained.

The theorem below characterizes the Bayes risk of the IRW policy in terms of both $M$ and $c$. 
\begin{theorem}\label{thm_multi}
Assume that there exists a constant $\delta>0$ independent of $M$ such that $D(h_l^{(d+k)}\|h_l^{(d)})>\delta$ for all $l=1,2,\ldots,\log_2 M$, $d=0,1,\ldots \min \lbrace L, 2^l \rbrace$, and $-d\leq k\leq\min\lbrace L, 2^l\rbrace -d$. For all $M$, $c$, and $L$ , we have
\begin{equation}\label{multi_risk}
\begin{split}
R\left({\Gamma_{\mbox{\footnotesize IRW}}}\right) \leq cLB\log_2 M +\frac{cL\log \frac{\log_2 M}{c}}{D(g_0\|f_0)} + O(c^2\log_2 M)+O(c),
\end{split}
\end{equation}
where $B$ is a constant independent of $c$, $M$, and $L$. Furthermore, the Bayes risk of IRW is order optimal in~$M$ for all $c$ and asymptotically optimal in $c$ for all $M$ sufficiently large.
\end{theorem}

\begin{figure}[b!]
   \centering
      \includegraphics[scale=0.4]{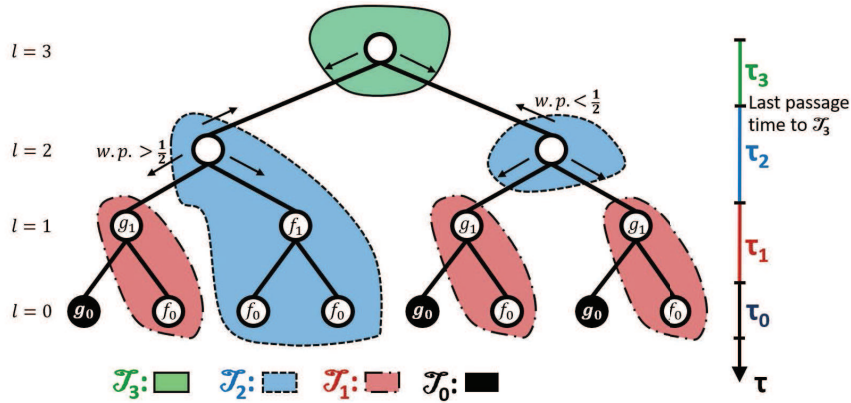}
      \caption{Partition of the tree into $\log_2 M +1$ subsets when there are multiple targets.}
      \label{fig:multi_tree_exit}
   \end{figure}

\begin{IEEEproof}
We present below the basic structure of the proof. Details can be found in Appendix~\ref{appx:proof_thm4}. Similar to the single-target case, the risk associated with a wrong decision is bounded by $O(c)$. In analyzing the sample complexity, a uniform bound on the sample complexity of finding each target, i.e., a uniform bound on each run of the random walk, is derived. Such a bound is again obtained by partitioning the tree into $\log_2 M+1$ subsets. These subsets, however, differ from the sub-trees in the single-target case. As illustrated in Fig.~\ref{fig:multi_tree_exit} for an example with $M=8$ and $L=3$, subset $\mathscr{T}_{0}$ consists of all the targets. Subset $\mathscr{T}_{l}$ ($l=1,\ldots, \log_2 M$) is the union of all the nodes on level $l$ that contain at least one target, and their entire left or right subtree if the subtree has no target. We then show that the successively defined last passage time to each of the subsets from $\mathscr{T}_{\log_2 M}$ to $\mathscr{T}_{1}$ are bounded by a constant. 
\end{IEEEproof}

\subsection{IRW Policy for Unknown $L$}
\label{ssec:unknown_targets}

When the number $L$ (may be zero) of targets is an unknown constant independent of $M$, the IRW policy can be augmented with a \emph{terminating phase} carried out at the root of the tree. Specifically, at the beginning of each run of the random walk when the local test takes samples from each child of the root node, if the local test indicates neither child contains undeclared targets, the policy enters the \emph{terminating phase} and starts taking samples from the root node itself. The SLLR of the root node is updated sequentially with each sample. A positive SLLR initiates the next run of the random walk. A negative SLLR that drops below $\log c$ terminates the detection process.

Following the similar lines of arguments in the proof of Theorem~\ref{thm_multi}, the Bayes risk of the IRW policy when the number of targets is unknown can be upper bounded by
\begin{equation}\label{unknown_risk}
\begin{split}
R\left(\Gamma_{\mbox{\footnotesize IRW}}\right) \leq cLB\log_2 M +\frac{cL\log \frac{\log_2 M}{c}}{D(g_0\|f_0)}+ \frac{c\log\frac{1}{c}}{D\left(h_{l_r}^{(L)}\Big\|h_{l_r}^{(L+1)}\right)}+ O(c^2\log_2 M)+O(c).
\end{split}
\end{equation}

\section{Discussions}
\label{sec:connections}

Several problems studied in different context can be cast as an active search problem. In this section, we discuss their connections and the applicability of the IRW strategy.

\subsection{Adaptive Sampling with Noisy Response}

Consider the problem of estimating a step function in $[0,1]$. Let $z^*\in [0,1]$ denote the unknown step of the function. The learner sequentially chooses sampling points in the interval and observes a noisy version of their values. Two commonly used noise models are the additive Gaussian noise model and the boolean noise model. In the former, the observations are Gaussian with mean $0$ or $1$, depending on whether the sampling point is to the left or the right of $z^*$. In the latter, the observations are Bernoulli random variables with parameters depending on the relative locations of the sampling points and $z^*$. The objective is to locate $z^*$ in a $\delta$-length interval with a probability no smaller than $1-\varepsilon$. This problem arises in active learning of binary threshold classifiers \cite{castro2008minimax} and stochastic root finding \cite{frazier2016probabilistic}.

The problem can be cast as one studied in this work. We partition the $[0,1]$ interval into $\delta$-length sub-intervals, which form the $M=1/\delta$ leaf nodes with the sub-interval containing $z^*$ being the target. Successively combining two adjacent sub-intervals leads to a binary tree with the root being the entire interval of $[0,1]$. What remains to be specified is the local test on a node. Since each node on the tree is a sub-interval of $[0,1]$, there remains the issue of which points in this sub-interval to sample when we prob this node. One way to translate the problem is to set the sampling points to the boundaries of each sub-interval. Thus, for the ternary hypothesis testing problem at each upper-level node, the observations are random vectors of dimension $4$, corresponding to sampling the two boundaries of each of the children. Similarly, at a leaf node, the observations are of dimension $2$. The local tests can be easily extended.

Most of the existing work on adaptive sampling is based on a Bayesian approach with a binary noise model. A popular Bayesian strategy, the probabilistic bisection algorithm, which updates the posterior distribution of the step location after each sample (based on the known model of the noisy response) and chooses the next sampling to be the median point of the posterior distribution. Several variations of the method have been extensively studied in the literature~\cite{waeber2013bisection, castro2008active,Tara_ISIT17,chiu2016sequential,lalitha2017improved}. In particular, in~\cite{lalitha2017improved}, Lalitha et al. proposed a two-stage algorithm based on the posterior matching method and showed the gain in sample complexity over non-adaptive/open-loop strategies. However, the update and sorting of the posterior probabilities at each sample require $O(M\log M$) computation and memory complexity. In contrast, the IRW approach assumes no prior distribution and has $O(1)$ computation and memory complexity. This is made possible by effectively localizing data processing to small subsets of the input domain based on the tree structure, which also allows dynamic allocation of limited data storage.

\subsection{Noisy Group Testing and Compressed Sensing}

In the group testing problem, the objective is to identify the defective items in a large population by performing tests on subsets of items that reveal whether the tested group contains any defective items (classic Boolean group testing) or the number of defective items in the tested group (quantitative group testing). Similar to the group testing problem, objective of the compressed sensing~\cite{atia2012boolean} is to recover a sparse single with aggregated observations.

The above various formulations of group testing and the compressed sensing problems can be mapped to an active search problem with specific observation models (e.g., Bernoulli distribution for noisy Boolean group testing, sum-observation model for the quantitative and threshold group testing). The individual items (signal components) in the group testing (compressed sensing) problems are mapped to the leaf nodes of a tree. The action of testing a node on the tree corresponds to a group test.

Most existing work on Boolean group testing assumes error-free test outcomes. There are several recent studies on the noisy group testing that assume the presence of one-sided noise~\cite{atia2012boolean, Tan&Atia:2014} or the symmetric case with equal size-independent false alarm and miss detection probabilities\cite{Cai&etal:2013, Chan&etal:2014}. In some extended group testing models such as the noisy quantitative group testing~\cite{QGT_TSP18} and threshold group testing~\cite{TGT}, the issue of sample complexity in terms of the detection accuracy is absent in the basic formulation. Most of the existing results on noisy group testing as well as compressed sensing focus on non-adaptive open-loop strategies that determine all actions in one shot \emph{a priori}. The disadvantages of non-adaptive test plans lie in the computational complexity of the coding/decoding processes and high storage requirement.

The proposed IRW policy provides an \emph{adaptive} solution to solve the group testing and compressed sensing with little offline or online computation and low memory requirement. The policy works for the general noisy observation models. Although the adaptive group testing strategies do not necessarily conform to a predetermined tree structure, the IRW policy offers asymptotic optimality in both population size and reliability constraint.

\subsection{Channel Coding with Feedback}

In channel coding with feedback~\cite{burnashev1976data, horstein1963sequential}, the encoder transmits symbols adaptively based on the receiving history of the decoder due to the availability of a noiseless feedback channel. The decoder then uses all the received symbols to decode the message.

The coding problems over any stationary discrete-input memoryless channel can be mapped to the problem studied in this work. Without loss of generality, consider a Discrete Memoryless Channel (DMC) where the output is also discrete. Let $\lbrace I_1, I_2, \ldots, I_J\rbrace$ and $\lbrace O_1, O_2, \ldots, O_K \rbrace$ denote, respectively, the input and output alphabets. Let $\mathbf{P}= \lbrace p_{j,k}\rbrace$ where $j=1,\ldots,J$, $k=1,\ldots, K$ be the channel transition probability matrix. Let $M$ be the number of messages. The objective is a coding scheme that transits these $M$ messages successfully with probability no smaller than~$1-\varepsilon$.

The above coding problem can be mapped to an active search problem on a binary tree with $M$ leaf nodes of which one is the target, i.e., the message being transmitted. The binary splitting structure generates a representation of the location of the target with a $\lbrace 0,1\rbrace$ codeword of which the length equals $\log_2 M$ (i.e., for each node, the left branch represents $0$ and the right branch represents $1$). The test on each level of the tree corresponds to sending the next bit or correcting the previous bit of the source code. The distribution of the target $g_0$ is set as the probability mass vector $\lbrace p_{j_g^*, k}\rbrace_{k=1}^{K}$ and the distribution of the non-target node $f_0$ is set as the probability mass vector $\lbrace p_{j_f^*, k}\rbrace_{k=1}^{K}$, where $j_g^*$ and $j_f^*$ are the two most distinguishable symbols transmitted through the channel and are defined as
\begin{equation}\label{DMC_map}
\left(j_g^*, j_f^*\right) = \arg\max_{\left(j_g,j_f\right)} \sum_{k=1}^{K} p_{j_g,k} \log \frac{p_{j_g,k}}{p_{j_f,k}}, \forall j_g, j_f = 1,\ldots, J.
\end{equation}
The observation distribution of a node on level $l\geq 1$ also follows $g_0$ if it contains the target or $f_0$ if it does not. It is not difficult to see that the action of sampling a node which contains the target in the target search problem corresponds to the action of sending $j_g^*$ through the channel, and the action of sampling a node which does not contain the target corresponds to the action of sending $j_f^*$ through the channel. The corresponding observations at the receiving end of the channel would following $g_0$ or $f_0$, respectively. 

The proposed IRW policy can be mapped to a coding scheme for the transmission problem. If the next bit of the source code is $0$, i.e., the left child of the current node contains the target, the sender sends symbol $j_g^*$ $K_l$ times following by sending symbol $j_f^*$ $K_l$ times through the channel. If the next bit of the source code is $1$, i.e., the right child contains the target, the sender sends symbol $j_f^*$ $K_l$ times following by sending symbol $j_g^*$ $K_l$ times through the channel. If there is an error in the transmission of previous bits, i.e., neither of the children contains the target, the sender sends symbol $j_f^*$ $2K_l$ times to inform the encoder to correct the previous bit. After each local test ($2K_l$ times channel usages), a bit of the source code is decoded correctly with probability greater than $\frac{1}{2}$. If a bit is decoded incorrectly, it would be revisited and corrected later with probability greater than $\frac{1}{2}$. When the random walk arrives at a leaf node, i.e., the full codeword has been transmitted and decoded, the sender keeps sending symbol $j_g^*$ if the entire codeword has been decoded correctly until the log-likelihood ratio at the receiver is large enough. If any bit of the codeword is decoded incorrectly, the sender keeps sending $j_f^*$. This step of sending the confirmation bits corresponds to the local test on a leaf in the IRW policy. \par

A similar connection between the active search problem and channel coding with feedback was discusses in~\cite{lalitha2017improved}, where an additive Gaussian noise channel with binary input was considered. The IRW strategy applies to more general channel models. Its advantage in computation and memory efficiency as discussed in the case of adaptive sampling applies here as well.

\section{Simulation Examples}
\label{sec:simulation}

We now provide numerical examples comparing the performance of the IRW policy with that of the Chernoff test and the DGF test developed in~\cite{Cohen2015active}.

Consider the problem of detecting $L$ heavy hitters among Poisson flows where the measurements are exponentially-distributed packet inter-arrival times. For the leaf-node, $g_0$ and $f_0$ are exponential distributions with parameters $\lambda_g$ and $\lambda_f$, respectively. The aggregated flows follow the corresponding exponential distributions with the parameters equal to the sum of the parameters of their children at the leaf level.\par

Under the same action space, which is given by all nodes on the tree, it can be shown that the resulting Chernoff test probes only the leaf nodes. Specifically, at each time $t$, all the leaf nodes are sorted based on their SLLRs. If
\begin{equation}\label{DGF_cond1}
D(g_0\|f_0)/L \geq D(f_0\|g_0)/(M-L),
\end{equation}
the Chernoff test randomly and uniformly selects one node from the ones with the largest SLLR to the $L$th largest SLLR; 
if
\begin{equation}\label{DGF_cond2}
D(g_0\|f_0)/L < D(f_0\|g_0)/(M-L),
\end{equation}
the Chernoff test randomly and uniformly selects one node from the ones with the $(L+1)$th largest SLLR to the smallest SLLR. The Chernoff and the IRW tests have the same stopping and decision rules.

   \begin{figure}[b!]
      \centering
      \includegraphics[scale=0.46]{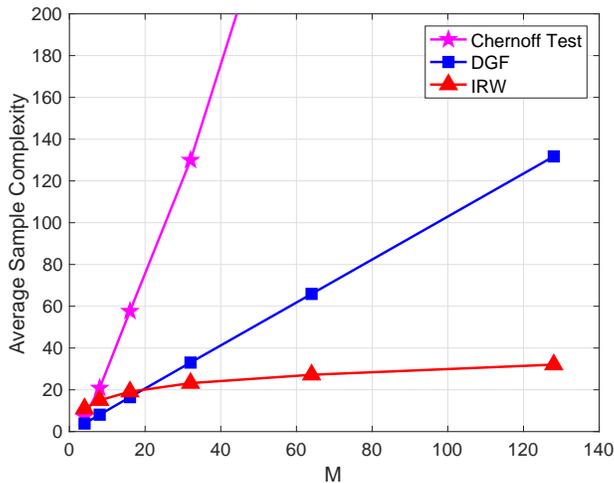}
      \caption{Performance comparison ($L=1$, $\lambda_g=10, \lambda_f=0.01, K_l=3, c=10^{-13}$, $M=4, 8, \ldots, 128.$)}
      \label{fig:compare3_single}
   \end{figure}

   \begin{figure}[t!]
      \centering
      \includegraphics[scale=0.46]{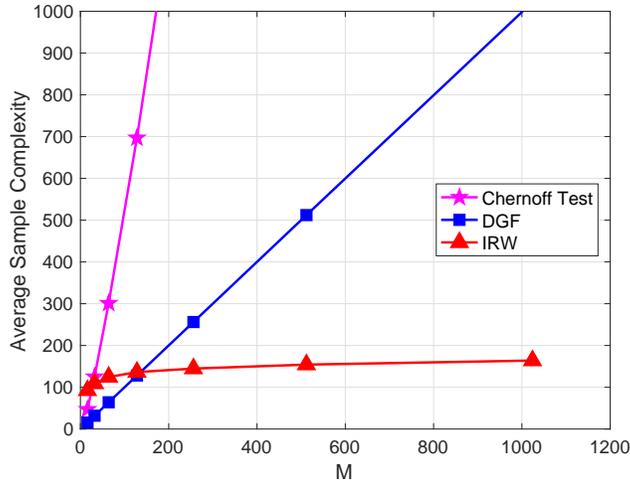}
      \caption{Performance comparison ($L=5$, $\lambda_g=10, \lambda_f=0.001, c=5\times 10^{-5}$, $M=16, 32, \ldots, 1024.$)}
      \label{fig:compare3_multi}
   \end{figure}
   \begin{figure}[t!]
      \centering
      \includegraphics[scale=0.46]{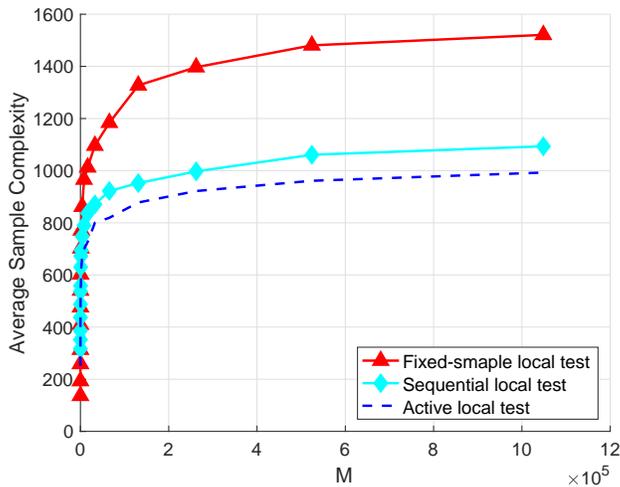}
      \caption{Performance comparison of IRW with different local tests. ($K_l=7$; thresholds for the sequential local test: $\gamma_1=1.0986$, $\gamma_0=-1.0986$; thresholds for the active local test: $\nu_1 = 0.9445$, $\nu_0 = -0.9445$; 1000 Monte Carlo runs.)}
      \label{fig:compare_local_test}
   \end{figure}

Fig.~\ref{fig:compare3_single} and Fig.~\ref{fig:compare3_multi} show simulation results on the sample complexity as a function of $M$ for $L=1$ and $L=5$, respectively. We observe that the Chernoff test has a sample complexity that scales linearly with $M$, while IRW offers a logarithmic order. The improvement at $M=20$ is already three-fold. 

Also shown in these figures is the DGF policy developed in~\cite{Cohen2015active}. The comparison with DGF is not on equal footing, since it was developed without assuming a tree structure. The DGF policy, as a deterministic policy, has a much smaller linear slope than the Chernoff test. The comparison between DGF and IRW is to show that by exploiting the hierarchical structure of the search space, more significant gain can be achieved in addition to efficient design of deterministic selection rules.

Next, we study the impact of different local tests on the finite-time performance of IRW. In the example shown in Fig.~\ref{fig:compare_local_test}, $L=1$, and the target-present and target-absent distributions are level independent and Bernoulli with parameters of $0.6$ and $0.4$, respectively. The confidence level of the local test is set to $p=0.5625$, which determines the number $K_l$ of samples in the fixed-sample-size test and the thresholds of the sequential and active tests. The improvement offered by the sequential and active local tests is evident. In addition, instead of designing $K_l$ for each level, one pair of thresholds for the SLLR that guarantee the biased global random walk can be used on all the higher level nodes.

\section{Conclusion}
\label{sec:conclusion}
In the paper, we considered the problem of detecting a few anomalous processes among a large number of processes under a tree observation model. The proposed active inference strategy induces a biased random walk on the tree and offers order optimality with the search space size and asymptotic optimality in terms of the reliability constraint. The proposed strategy is also efficient in terms of computation and memory requirement. By effectively localizing the data processing to small subsets of the search space, it has $O(1)$ computation and memory complexity.


\appendices

\renewcommand{\thesubsection}{\arabic{subsection}}

\section{Alternative Local Tests}
\label{appx:seq_local}

\subsection{Sequential Local Test}

A sequential version of the local test is to carry out SPRT on each child, one at a time. If the SPRT on the left child indicates it contains the target, the local test declare $H_1$. Otherwise, carry out SPRT on the right child and declare $H_2$ or $H_0$ based on the outcome of this SPRT.

We now specify the thresholds to be used in each SPRT. To ensure the required confidence level $p$ of the local test, the false alarm and miss detection probabilities of each SPRT should satisfy $(1-P_{\text{FA}})^2 > p$ and $(P_\text{FA})(1-P_\text{MD}) >p$. This leads to the following positive and negative thresholds, $\gamma_1$ and $\gamma_0$ respectively, of the SPRT
\begin{equation}\label{SPRT_thred}
\gamma_1 = \log \frac{1-P_{\text{MD}}}{P_{\text{FA}}}, \quad \gamma_0 = \log \frac{P_{\text{MD}}}{1-P_{\text{FA}}}.
\end{equation}

%
%

\subsection{Active Local Test}

We now present a fully active test that adaptively determines which child to sample. 

The selection rule is to sample the child with a greater SLLR at each time. The stopping rule is determined by a pair of thresholds $\nu_0$ and $\nu_1$. The test stops as soon as $\max\left\lbrace S_L, S_R\right\rbrace \leq \nu_0$, and declares $H_0$; or when $\max\left\lbrace S_L, S_R\right\rbrace \geq \nu_1$, in which case either $H_1$ or $H_2$ is declared, depending on which $S_L$ or $S_R$ exceeds $\nu_1$. The thresholds are set as following:
\begin{equation}\label{active_thresholds}
\nu_1 = \log \frac{P_{11}}{P_{01}}, \quad \nu_0 = \log \frac{P_{10}}{P_{00}},
\end{equation}
where $P_{11}$ denote the probability of declaring hypothesis $H_{1}$ when $H_{1}$ is true, and $P_{00}$ are defined similarly. To ensure a confidence level of $p$, we set $P_{00} = P_{11} = p$, $P_{01}=(1-P_{00})/2$ and $P_{10}= (1-P_{11})/2$.\par

We now show the derivation of the active local test. Let $X_1, X_2, \ldots, ...$ denote the samples taken from the left node and $Y_1, Y_2, \ldots, ...$ denote the samples taken from the right node. Let $p_1$ and $p_0$ denote the distributions of the samples taken from the node that contains or not contains the target, respectively. 
Define
\begin{eqnarray}
&& \Lambda_x^k := \prod_{i=1}^{k} \frac{p_1 (X_i)}{p_0 (X_i)},\ k = 1,2,\ldots\\
&& \Lambda_y^l := \prod_{j=1}^{l} \frac{p_1 (Y_i)}{p_0 (Y_i)},\ l = 1,2,\ldots
\end{eqnarray}
which are the likelihood ratios (LRs) of the left and right child nodes, respectively. The log-likelihood ratios (LLRs) of the two nodes are given by:
\begin{eqnarray}
&& S_x^k := \log \Lambda_x^k = \sum_{i=1}^{k} \log\frac{p_1 (X_i)}{p_0 (X_i)},\ k = 1,2,\ldots\\
&& S_y^l := \log \Lambda_y^l = \sum_{i=1}^{l}  \log\frac{p_1 (Y_i)}{p_0 (Y_i)},\ l = 1,2,\ldots
\end{eqnarray}

To simplify the notation, let $x:=(x_1,\ldots, x_k)$, $y:=(y_1,\ldots, y_l)$, and write $p_j(x) = \prod_{i=1}^k  p_j(x_i)$ and $p_j(y) = \prod_{i=1}^l  p_j(y_i)$, $j=0,1$. 

The decision sets of the active local test can be written as:
\begin{eqnarray}
&& R_0 :=\left\lbrace x, y: \Lambda_x^k \leq \sigma_0,\ \Lambda_y^l \leq \sigma_0 \right\rbrace,\\
&& R_1 :=\left\lbrace x, y: \Lambda_x^k \geq \sigma_1,\ \Lambda_y^l \leq 1 \right\rbrace,\\
&& R_2 :=\left\lbrace x, y: \Lambda_x^k \leq 1,\ \Lambda_y^l \geq \sigma_1 \right\rbrace.
\end{eqnarray}
$P_{11}$ can be written in terms of the decision set $R_1$ as follows:
\begin{equation}
\begin{split}
P_{11} &= \int_{R_1} p_1(x) p_0(y) dx dy\\
&= \int_{R_1} \frac{p_1(x)}{ p_0(x)}  p_0(x) p_0(y)dxdy\\
&= \int_{R_1} \Lambda_x^k  p_0(x) p_0(y)dxdy\\
&\geq \sigma_1 \int_{R_1}  p_0(x) p_0(y)dxdy\\
&= \sigma_1 P_{01}. \\
\end{split}
\end{equation}
$P_{00}$ can be written in terms of the decision set $R_0$ as
\begin{equation}
\begin{split}
P_{00} &= \int_{R_0} p_0(x) p_0(y) dx dy\\
&=\int_{R_0} \frac{p_0(x)}{ p_1(x)}  p_1(x) p_0(y)dxdy\\
&= \int_{R_0} \frac{1}{\Lambda_x^k}  p_1(x) p_0(y)dxdy\\
&\geq \frac{1}{\sigma_0}\int_{R_0}  p_1(x) p_0(y)dxdy\\
&= \frac{1}{\sigma_0} P_{10}.
\end{split}
\end{equation}
Similarly, we can get $P_{22}\geq \sigma_1 P_{02}$ and $P_{00} \geq \frac{1}{\sigma_0}P_{20}$. In the detection on each level, it is common to set that $P_{11}= P_{22}$, $P_{10}= P_{20}$ and $P_{01}=P_{02}$.
These expressions give us bounds on the thresholds which are necessary to achieve $P_{11}$, $P_{10}$, $P_{00}$, and $P_{01}$:
\begin{eqnarray}
&& \sigma_1 \leq \frac{P_{11}}{P_{01}},\\
&& \sigma_0 \geq \frac{P_{10}}{P_{00}}.
\end{eqnarray}
We then set 
\begin{equation}
\sigma_1 = \frac{P_{11}}{P_{01}},\quad \sigma_0 = \frac{P_{10}}{P_{00}}.
\end{equation}
It is easy to see that $P_{11} + P_{10} + P_{12} = 1$, and $P_{00}+P_{01} +P_{02} = P_{00}+2P_{01}=1$.  In the RWT policy, we require $P_{11}>\frac{1}{2}$ and $P_{00}>\frac{1}{2}$. It can be seen that $\sigma_1>1$ and $\sigma_0<1$. When changing to the log-likelihood ratio, we have $\nu_1 = \log \sigma_1 = \log \frac{P_{11}}{P_{01}}$, and $\nu_0 = \log \sigma_0 =\log \frac{P_{10}}{P_{00}}$.

When using the IRW policy, we update the SLLRs after taking each sample. The test keeps sampling if 
$$\log \frac{P_{10}}{P_{00}}<\max\left\lbrace S_x^k, S_y^l\right\rbrace < \log \frac{P_{11}}{P_{01}}.$$
With the proposed active local test, we can also guarantee that $P_{12} \leq P_{02}$, since it is not difficult to see that with the decision set $R_2$, $P_{12}$ can be written as
\begin{equation}
\begin{split}
P&_{12} = \int_{R_2} p_1(x) p_0(y) dx dy = \int_{R_2} \frac{p_1(x)}{ p_0(x)}  p_0(x) p_0(y)dxdy\\
&= \int_{R_2} \Lambda_x^k  p_0(x) p_0(y)dxdy \leq \int_{R_2}  p_0(x) p_0(y)dxdy = P_{02}.
\end{split}
\end{equation}
After setting $P_{01}= P_{02}$, we have $P_{12}\leq P_{01}$.

\section{Proof of Theorem~\ref{main_thm}}
\label{appx:proof_thm1}
Without loss of generality (due to the symmetry of the binary tree structure), we assume that the left-most leaf is the target. We focus on the IRW with fixed-sample local tests. \par

The random walk on the tree can be divided into two states. The first state is the random walk on upper-level nodes of the binary tree. In this state, at each time, after taking $K_l$ samples, we either zoom-in to one child node or zoom-out to the parent node. As a result, the distance between the current node to the target is defined as the sum of the discrete distance to the target node on the tree and the threshold $\log \frac{\log_2 M}{c}$, which either decreases by one when zooming-in or increases by one when zooming-out after tanking $2K_l$ samples from the children. Once arriving at a leaf node, the test arrives at the second state, where samples are taken one by one from the current node until the cumulative SLLR exceeds the threshold or becomes negative. The cumulative SLLR can be viewed as a discrete time random walk with random continuous step size which is the LLR of each sample. For all the non-target leaf-nodes, we define the distance between the node to the target as the sum of the discrete distance on the tree, the cumulative SLLR of the current node, and the threshold. For the target node, we define the distance to the target as the difference between the threshold and the current cumulative SLLR of the target node. During the search process, these two different states happen consecutively in Phase~I of the IRW policy. \par

Let $W_n$ denote the random variable of the step size of the random walk at time~$n$. When the IRW is in the first state (i.e., random walk on high-level nodes), depending on the current level $l>0$, $W_n$ has the following distribution:
\begin{equation}
\Pr (W_n) = \begin{cases}
p_l^{(g)}, & \text{for\ } W_n = -1,\cr
1-p_l^{(g)}, & \text{for\ } W_n = 1,
\end{cases}
\end{equation}
if the node is located at a sub-tree that contains the target, or
\begin{equation}
\Pr (W_n) = \begin{cases}
p_l^{(f)}, & \text{for\ } W_n = -1,\cr
1-p_l^{(f)}, & \text{for\ } W_n = 1,
\end{cases}
\end{equation}
if the node is located at a sub-tree that does not contain the target. Since $p_l^{(g)}>\frac{1}{2}$ and $p_l^{(f)}>\frac{1}{2}$ for all $l=1,2,\ldots, \log_2 M$, we have:
$$\mathbb{E}[W_n] = 1-2p_l^{(g)} \text{\ or\ } 1-2p_l^{(f)},$$
which are both less than $0$. \par

For the second state, let $Y_0$ and $Z_0$ denote the random variables with distributions $g_0$ and $f_0$, respectively. The LLR will be either $-\log\frac{g_0(Y_0)}{f_0(Y_0)}$ or $\log\frac{g_0(Z_0)}{f_0(Z_0)}$. It is not difficult to see that for the target node, we have:
\begin{equation}\label{ew_target}
\mathbb{E}[W_n] = \mathbb{E}\left[-\log\frac{g_0(Y_0)}{f_0(Y_0)}\right]=-D(g_0\|f_0)<0,
\end{equation}
and for all the non-target nodes, we have:
\begin{equation}\label{ew_normal}
\mathbb{E}[W_n] = \mathbb{E}\left[\log\frac{g_0(Z_0)}{f_0(Z_0)}\right] = -D(f_0\|g_0)<0.
\end{equation}

We further assume that the distribution of $-\log\frac{g_0(Y_0)}{f_0(Y_0)}$ and $\log\frac{g_0(Z_0)}{f_0(Z_0)}$ are light-tailed distributions\footnote{A random variable $X$ with the cumulative distribution function $F(x)$ is light-tailed if and only if $\int_\mathbb{R} e^{\lambda x}dF(x)<\infty$ for some $\lambda>0$~\cite{foss2011introduction}.}.

The following lemma characterizes the distributions of the last passage times $\tau_i$ as defined in Section~\ref{ssec:main_structure}.
\begin{lemma}\label{tau_exp_lemma}
For all $\tau_i$ with $i=1,\ldots, \log_2 M$, there exist $\alpha>0$ and $\gamma>0$ which are independent of $M$ and $c$, such that
\begin{equation}
\Pr (\tau_i \geq n)\leq \alpha e^{-\gamma  n},\ \forall n\geq 0.
\end{equation}
\end{lemma}
\begin{IEEEproof}
We first prove this lemma for $\tau_{\log_2 M}$ which is the last passage time of the sub-tree at the root that does not contain the target. \par

Let $S_t$ denote the distance to the target at time $t$. The IRW policy starts at the root node. Therefore the initial distance to the target is $S_0 = \log_2 M +\log \frac{\log_2 M}{c}$. Define
\begin{equation}
\tau^* = \sup \left\lbrace t\geq 0: S_t\geq S_0 \right\rbrace
\end{equation}
as the last time when the search approaches the distance to the target which is greater than $S_0$. It is not difficult to see that
\begin{equation}
\tau_{\log_2 M} \leq \tau^*.
\end{equation}
Therefore, we have
\begin{equation}\label{tau_tau*}
\Pr(\tau_{\log_2 M} \geq n) \leq \Pr(\tau^* \geq n).
\end{equation}
Based on the definition of $\tau^*$, we have
\begin{equation}\label{tau*ineq}
\begin{split}
\Pr(\tau^* >n) = \Pr \left(\sup \lbrace t\geq 0: S_t\geq S_0\rbrace > n\right)\leq \sum_{t=n}^{\infty} \Pr (S_t \geq S_0) = \sum_{t=n}^{\infty} \Pr \left( \sum_{j=1}^{t} W_j \geq 0 \right).
\end{split}
\end{equation}

Let $\mu_j$ denote the mean value for each $W_j$, where $\mu_j<0$ for all $j=1,2,\ldots,t$.
By applying the Chernoff bound to the sum of independent random variables $\displaystyle\sum_{j=1}^t W_j$, for all $s>0$ we have:
\begin{equation}\label{chernoff_bound}
\Pr \left( \sum_{j=1}^{t} W_j \geq 0 \right)\leq \mathbb{E}\left[ e^{s\sum_{j=1}^t W_j }   \right] = \prod_{j=1}^{t} \mathbb{E}\left[e^{s W_j} \right]. 
\end{equation}
Note that the moment generating function (MGF) of each $W_j$ is equal to one at $s=0$. Furthermore, since $\mathbb{E}\left[W_j\right]<0$ is strictly negative for all $j\geq 1$, differentiating the MGFs of all $W_j$ with respect to $s$ yields strictly negative derivatives at $s=0$. Because all $W_j$'s are light-tailed distributions for all possible distributions of $W_j$, there exist $s>0$ and $\gamma>0$ such that $\mathbb{E}\left[e^{s W_j}\right]$ is strictly less than $e^{-\gamma}<1$. Hence, from~\eqref{chernoff_bound}, we have
\begin{equation}
\Pr\left( \sum_{j=1}^{t} W_j \geq 0 \right) \leq e^{-\gamma t}.
\end{equation}
Due to~\eqref{tau*ineq}, we have
\begin{equation}
\begin{split}
\Pr(\tau^* >n) &\leq \sum_{t=n}^{\infty} \Pr \left( \sum_{i=1}^{t} W_i \geq 0 \right)\leq \sum_{t=n}^{\infty} e^{-\gamma t}= \frac{e^{-\gamma n}}{1-e^{-\gamma}}.
\end{split}
\end{equation}
Let $\alpha = \frac{1}{1-e^{-\gamma}}$, with \eqref{tau_tau*}, we complete the proof of Lemma~\ref{tau_exp_lemma} proved for $\tau_{\log_2 M}$. Due to the recursive definitions of $\tau_1, \tau_2,\ldots, \tau_{\log_2 M}$, the proof follows the same procedure for all other $\tau_i$.
\end{IEEEproof}

Based on Lemma~\ref{tau_exp_lemma}, we get the following lemma that characterizes the expected value of $\tau_i$.

\begin{lemma}\label{lemma:tau_ave}
For all $\tau_i$ with $i=1,\ldots, \log_2 M$, there exists a constant $\beta>0$, such that
\begin{equation}
\mathbb{E}[\tau_i] \leq \beta.
\end{equation}
\end{lemma}
\begin{IEEEproof}
Based on the the tail-sum formula of expectation of non-negative random variables, we have
\begin{equation}
\begin{split}
\mathbb{E}[\tau_i] = \sum_{n=0}^{\infty} \Pr  [\tau_i>n]\leq \sum_{n=0}^{\infty} \alpha e^{-\gamma n} = \frac{\alpha}{1-e^{-\gamma}}=\frac{1}{(1-e^{-\gamma})^2} =\beta.
\end{split}
\end{equation}
\end{IEEEproof}

Now we are ready to prove Theorem~\ref{main_thm}. Based on Lemma~\ref{lemma:tau_ave}, it is not difficult to show that
\begin{equation}\label{eq_beta}
\mathbb{E}[\tau] \leq 2K_{\max} \sum_{i=1}^{\log_2 M} \mathbb{E}[\tau_i] +\mathbb{E}[\tau_0] \leq 2\beta K_{\max}\log_2 M +\mathbb{E}[\tau_0].
\end{equation}
When the observations are informative at all levels, $K_{\max}$ is bounded by a constant. As a result, the first term on the RHS of~\eqref{eq_beta} is upper bounded by $B\log_2 M$, where $B$ is a constant greater than $2\beta K_{\max}$.

For the last stage, $\tau_{0}$ is a stopping time with respect to the i.i.d. sequence of the LLR $\left\lbrace \log\frac{g_0(X_n)}{f_0(X_n)}: n\geq 1 \right\rbrace$, where $X_n$ denotes an i.i.d. random variable with distribution $g_0$.

Due to the Wald's Equation\cite{wald1945}, we have
\begin{equation}
\mathbb{E}\left[ \sum_{n=1}^{\tau_0} \log\frac{g_0(X_n)}{f_0(X_n)}\right] = \mathbb{E}[\tau_0] \mathbb{E}\left[  \log\frac{g_0(X_n)}{f_0(X_n)} \right].
\end{equation}
i.e.,
\begin{equation}
\log \frac{\log_2 M}{c}+ R_b =  \mathbb{E}[\tau_0] D(g_0\|f_0),
\end{equation}
where $R_b$ is the overshooting at the threshold. Due to Lorden's inequality~\cite{lorden1970excess}, we have
\begin{equation}
\mathbb{E}[R_b] \leq \frac{\mathbb{E}\left[\left(\log\frac{g_0(X_n)}{f_0(X_n)}\right)^2\right]}{\mathbb{E}\left[\log\frac{g_0(X_n)}{f_0(X_n)}\right]}.
\end{equation} 
Assuming that the first two moments of the LLR are finite, we then have
\begin{equation}\label{ex_tau0}
\mathbb{E}[\tau_0] = \frac{\log \frac{\log_2 M}{c}}{D(g_0\|f_0)} +O(1).
\end{equation}

The following lemma characterizes the error probability of the IRW policy. 
\begin{lemma}\label{lemma:error}
The error probability of the IRW policy is upper bounded by:
\begin{equation}
P_e \leq  \beta c = O(c) .
\end{equation}
\end{lemma}
\begin{IEEEproof}
When the IRW policy arrives a non-target node, say node $j$, the probability of error (accepting $H_j$) equals to $\Pr(S_j\geq \log\frac{\log_2 M}{c})$. For the seqnentual probability ratio test, Wald~\cite{wald1947} shows that 
\begin{equation}
\Pr (S_j \geq \log\frac{\log_2 M}{c}) \leq \exp\left[ -\log \frac{\log_2 M}{c}  \right] =\frac{c}{\log_2 M}.
\end{equation}

Let $N$ denote the random number of times of visiting these non-target leaf nodes in the IRW policy. The conditional error probability is upper bounded by $\frac{Nc}{\log_2 M}$. Based on the proof of Theorem~\ref{main_thm}, the expected value of $N$ is upper bounded by $\beta\log_2 M$. Therefore, by taking expectation, the error probability is bounded by
\begin{equation}
P_e \leq \frac{c}{\log_2 M} \cdot \mathbb{E}[N]\leq \frac{c}{\log_2 M}\cdot \beta\log_2 M=\beta c =O(c).
\end{equation}
\end{IEEEproof}

Combining~\eqref{eq_beta}, \eqref{ex_tau0} and Lemma~\ref{lemma:error} completes the proof.\par

\section{Proof of Theorem~\ref{poly} and Theorem~\ref{exp}} 
\label{appx:proof_thm23} 

Follow the same lines of argument in the proof of Theorem~\ref{main_thm}, the sample complexity of the last stage $\tau_0$ still satisfies~\eqref{ex_tau0}. We now give the upper bound of the sample complexity of the first $\log_2 M$ stages for the test with a fixed-size local test. \par

We focus on the Bernoulli distribution model, where $g_l$ and $f_l$ are Bernoulli distributions with false negative and false positive rates equal to $\mu_l$.
In order to get the relation between $K_l$ and $\mu_{l-1}$, we first introduce the following lemma. 
\begin{lemma}[{\cite{Michael_book_prob}}]
\label{K_bound}
Let $X_1, \ldots, X_n$ be independent Poisson trials such that $\Pr (X_i)=p_i$. Let $X=\sum_{i=1}^{n} X_i$ and $\nu = \mathbb{E}[X]$. Then, the following Chernoff bounds hold for $0< \delta\leq 1$,
\begin{eqnarray}
&& \Pr (X\geq (1+\delta)\nu) \leq e^{-\nu \delta^2 /3};\label{cher1}\\
&& \Pr (X\leq (1-\delta)\nu) \leq e^{-\nu \delta^2 /3}.\label{cher2}
\end{eqnarray}
\end{lemma}

The IRW policy requires that $p_l^{(g)}$ and $p_l^{f}$ will satisfy~\eqref{pg_pf}. For $p_l^{f}$, we need to find the value of $K_l$ such that 
$$\Pr\left(\sum_{n=1}^{K_l}  \log \frac{g_{l-1}(Z_n)}{f_{l-1}(Z_n)}<0\right)>\lambda.$$
i.e.,
$$\Pr\left(\sum_{n=1}^{K_l}  \log \frac{g_{l-1}(Z_n)}{f_{l-1}(Z_n)}\geq 0 \right)\leq 1-\lambda,$$
where $\lambda$ is a constant which satisfies $\frac{1}{2} \leq \lambda^2<1$. The true distribution of $Z_n$ is $f_{l-1}$ which is Bernoulli with success probability $\mu_{l-1}$. If $z_n = 1$, we have $\log \frac{g_{l-1}(Z_n)}{f_{l-1}(Z_n)} = \log\frac{1-\mu_{l-1}}{\mu_{l-1}}$; if $z_n = 0$, we have $\log \frac{g_{l-1}(Z_n)}{f_{l-1}(Z_n)} = -\log\frac{1-\mu_{l-1}}{\mu_{l-1}}$. Therefore, the above probability can be written as
\begin{equation*}
\begin{split}
\Pr\left(\sum_{n=1}^{K_l}  \log \frac{g_{l-1}(Z_n)}{f_{l-1}(Z_n)}\geq 0 \right) = \Pr\left(\log\frac{1-\mu_{l-1}}{\mu_{l-1}}\sum_{n=1}^{K_l} (2Z_n -1)\geq 0 \right) =  \Pr\left(\sum_{n=1}^{K_l} (2Z_n -1)\geq 0 \right).
\end{split}
\end{equation*}
The second equation is true because $\mu_{l-1}<\frac{1}{2}$ for all $l$. Therefore, we need to find $K_l$ such that  
$$\Pr\left(\sum_{n=1}^{K_l} (2Z_n -1)\geq 0 \right) =  \Pr\left(\sum_{n=1}^{K_l} Z_n \geq \frac{K_l}{2} \right) \leq 1-\lambda.$$
Notice that $Z_n$'s are Poisson trials, and Lemma 4 applies. When applying Lemma 4, we have $\nu = K_l\mu_{l-1}$, $(1+\delta)\nu = \frac{K_l}{2}$, which means $\delta = \frac{1-2\mu_{l-1}}{2\mu_{l-1}}$. Then, we have
$$\Pr\left(\sum_{n=1}^{K_l} Z_n \geq \frac{K_l}{2} \right) \leq e^{-\nu\delta^2/3}\leq 1-\lambda.$$ Substituting $\nu$ and $\delta$, we have
$$K_l\cdot \mu_{l-1}\cdot \frac{1}{3} \cdot\frac{(1-2\mu_{l-1})^2}{4\mu_{l-1}^2} \geq \log (1-\lambda)^{-1}.$$
i.e.,
$$K_l \geq \frac{12\mu_{l-1} \log (1-\lambda)^{-1}}{(1-2\mu_{l-1})^2}.$$
Similarly, by applying Lemma 4, in order to have $p_g>\frac{1}{2}$, we need $$K_l\geq \frac{12(1-\mu_{l-1}) \log (1-\eta)^{-1}}{(1-2\mu_{l-1})^2},$$
where $\eta$ and $\lambda$ can be any value in $(\frac{1}{\sqrt{2}}, 1)$ such that $\eta\cdot\lambda>\frac{1}{2}$ and $\lambda^2 > \frac{1}{2}$. 
In order to have $p_l^{(g)}$ and $p_l^{(f)}$ both satisfying \eqref{pg_pf}, we choose $K_l$ greater than
\begin{equation}
\max\left\lbrace \frac{12(1-\mu_{l-1}) \log (1-\eta)^{-1}}{(1-2\mu_{l-1})^2},  \frac{12\mu_{l-1} \log (1-\lambda)^{-1}}{(1-2\mu_{l-1})^2} \right\rbrace. 
\end{equation} 
Since $\mu_l<\frac{1}{2}$, w.l.o.g., we choose
\begin{equation}\label{max_kl}
K_l = \frac{12(1-\mu_{l-1}) \log (1-\eta)^{-1}}{(1-2\mu_{l-1})^2}. 
\end{equation}
It is not difficult to see that $K_l$ increases with $\mu_{l-1}$. 
For any stage $l$, when $l=1,2,\ldots, \log_2 M$, the sample complexity in this stage is upper bounded by $2K_l\cdot \mathbb{E}[\tau_l]$.
Due to Lemma~\ref{lemma:tau_ave}, the total sample complexity from Stage $1$ to Stage $\log_2 M$ is thus upper bounded by
\begin{equation}\label{sum_kl}
\mathbb{E}[\tau] \leq \sum_{l=1}^{\log_2 M}2K_l\cdot \mathbb{E}[\tau_l] \leq \sum_{l=1}^{\log_2 M}2\beta K_l.
\end{equation}
For Theorem~\ref{poly}, if $\mu_l = \frac{1}{2} - (\frac{1}{2}-\mu_0)\cdot (l+1)^{-\alpha}$, due to~\eqref{max_kl} and~\eqref{sum_kl}, we have
\begin{equation}
\mathbb{E}[\tau] \leq B' \sum_{l=1}^{\log_2 M} l^{2\alpha},
\end{equation}
where $B'= \frac{6\beta\log(1-\eta)^{-1}}{(\frac{1}{2}-\mu_0)^2}$ is a constant. By using the Faulhaber's formula, we have $$\sum_{l=1}^{\log_2 M} l^{2\alpha} = O((\log_2 M)^{2\alpha+1}).$$
Thus, Theorem~\ref{poly} is proved. \par

Similarly, for Theorem~\ref{exp}, if $\mu_l = \frac{1}{2} - (\frac{1}{2}-\mu_0)\cdot \alpha^{-l}$, we have
\begin{equation}\label{sum_geo}
\mathbb{E}[\tau] \leq B' \sum_{l=1}^{\log_2 M} \alpha^{2(l-1)}.
\end{equation}
By summing up the geometric terms in~\eqref{sum_geo}, we can show that 
\begin{equation}
\mathbb{E}[\tau] \leq \tilde{B} (\alpha^2)^{\log_2 M} = \tilde{B} M^{\frac{2}{\log_\alpha 2}},
\end{equation}
where $\tilde{B} = \frac{1}{\alpha^2-1}B'$. Thus, Theorem~\ref{exp} is proved.

\section{Proof of Theorem~\ref{thm_multi}}
\label{appx:proof_thm4}

To prove Theorem~\ref{thm_multi}, we first show that the sample complexity of the IRW policy satisfies
\begin{eqnarray}\label{multi_delay}
Q(\Gamma_{\mbox{\footnotesize IRW}})\leq LB\log_2 M +\frac{L\log \frac{\log_2 M}{c}}{D(g_0\|f_0)} + O(c\log_2 M).
\end{eqnarray}

In the proof of the single-target case, we have defined the random walk as the distance from the current testing node to the target. Since the random walk is biased, i.e., the IRW policy guarantees that the probability of approaching the target is always greater than $\frac{1}{2}$, the expectation of each step of the random walk tends to approach the target. By using the Chernoff bound, we have shown that the last passage time $\mathbb{E}[\tau_l]$ on the tree $\mathscr{T}_{l}$ for all $l=1,2,\ldots,\log_2 M$ is upper bounded by a constant. Then sample complexity in the first $\log_2 M$ stages thus has a logarithmic-order. For the last stage on $\mathscr{T}_0$, it can be shown that $\mathbb{E}[\tau_0] = \frac{\log \frac{\log_2 M}{c}}{D(g_0\|f_0)}+O(1)$. 

The basic idea to prove~\eqref{multi_delay} is similar to the one-target case. For multiple-target detection, we need to find a proper random variable that defines the increment of the random walk. A negative expected increment is desired so that the r.v. tends to zero as the IRW policy approaches the targets. Then, by using the Chernoff bound, we can get the similar upper bound of the detection delay for the multi-target case. 

The IRW policy is designed to find the targets one by one. As the process continuous, wrong declarations might propagate from the previous rounds. We consider the two cases for the search process on the tree with or without existing wrong declarations separately. 

\subsection{Search on the tree without existing wrong declarations}

Unlike the one-target case, we modify the definition of $\mathscr{T}_{l}$ for all $l=1,2,\ldots,\log_2 M$ for the multiple targets detection case. As illustrated in Fig.~\ref{fig:multi_tree_exit} for $M=8$ and $L=3$, our approach is to partition the tree into $\log_2 M+1$ disjoint sets of nodes. Similar to the one-target case, the detection process of finding any one of the targets is then partitioned into $\log_2 M +1$ stages by the successively defined last passage time to each of the set of nodes from the upper level to the lowest level.  

We start by finding the first target. The random walk on the tree has two states. The first state is the random walk on the upper-level nodes of the binary tree. Once arriving at a leaf node, the test moves to the second state, in which samples are taken one by one from the current node until the cumulative SLLR exceeds the threshold or becomes negative. Without loss of generality, we enumerate all the targets with index $1$ to $L$ from left to right.  For any node $v$ on the tree, we define $D_{\min}(v)$ as
\begin{equation}\label{multi_dist}
D_{\min}(v) := \min_{i=1, \ldots, L} \left\lbrace D_i(v)\right\rbrace,
\end{equation}
where $D_i(v)$ is the distance on the tree between the current node $v$ to the $i$th target.

For all the non-target leaf-nodes $v$, we define the distance between the node to the target as the sum of $D_{\min}(v)$, the cumulative SLLR of the current node, and the threshold $\log\frac{\log_2 M}{c}$. For the target node, we define the distance to the target as the difference between the threshold and the current cumulative SLLR of the target node.

Let $W_n$ denote one step of the global random walk at time $n$. When the IRW is in the first state, given the current node $v$, $W_n =\Delta D_{\min}(v)$ can be either $1$ or $-1$, which has the distribution
\begin{equation}
\begin{split}
\Pr \left(W_n\right) = \Pr \left(\Delta D_{\min}(v)\right) = 
\begin{cases}
p_l(v),  &\text{ for } W_n = \Delta D_{\min}(v)=-1,\cr
1-p_l(v), &\text{ for } W_n = \Delta D_{\min}(v)=1.
\end{cases}
\end{split}
\end{equation}

Under the IRW policy, for node $v$ on level $l$, after taking $K_l$ samples, the random walk has probability $p_l(v)>\frac{1}{2}$ to approach the targets in the tree rooted at the current node or $p_l(v)>\frac{1}{2}$ to zoom out of the current node if it contains no targets. Therefore, we have
$$\mathbb{E}[W_n] = \mathbb{E}[\Delta D_{\min}(v)] = 1-2p_l(v) <0,$$
which results in drifting toward at least one of the targets on the tree. 

In the second state, similar to the one-target case, we have~\eqref{ew_target} for the target nodes and~\eqref{ew_normal} for all the non-target leaves.

Similarly, let $\tau_i$ denote the last passage time to set $\mathscr{T}_i$. More specifically, $\tau_i$ is also the last time that the random walk has a distance greater or equal to $i+\log\frac{\log_2 M}{c}$ to all the targets. As a result, after $\tau_i$ has elapsed, the random walk will have a distance less than $i+\log\frac{\log_2 M}{c}$ to at least one of the targets. Then, using the same arguments as in the proof under the one-target case, we have that for all $\tau_i$, $i = 1,\ldots, \log_2 M$, there exists a constant $\beta > 0$, such that 
\begin{equation}
\mathbb{E}[\tau_i]\leq \beta.
\end{equation}

Therefore, the detection delay $\mathbb{E}[\tau]$ of finding a target in the first round is upper bounded by 
\begin{equation}\label{round_bound}
\mathbb{E}[\tau] \leq B\log_2 M + \frac{\log \frac{\log_2 M}{c}}{D(g_0\|f_0)} + O(1).
\end{equation}

Similar to the one-target case, the probability of making the first detection error in this round is bounded by 
$$P_e \leq \beta c = O(c).$$

For the subsequent $L-1$ rounds used for finding the remaining $L-1$ targets, as long as there are no detection errors, the detection delay of each round can be bounded by~\eqref{round_bound}. Similarly, the probability of making the first detection error in this round is upper bounded by $P_e \leq \beta c = O(c)$. Applying the union bound, we can find that, with probability at least $1-O(Lc)$, there would be no detection errors and the detection delay of finding all the $L$ targets is upper bounded by:
\begin{equation}\label{multi_bound}
\mathbb{E}[\tau_{\text{all}}] \leq LB\log_2 M + \frac{L\log \frac{\log_2 M}{c}}{D(g_0\|f_0)} + O(L).
\end{equation}

\subsection{Search on the tree with existing wrong declarations}

Assume that $L$ targets remained to be detected and there are $E$ detection errors. Due to the detection errors on the tree, the preference of the IRW policy to approach the targets may change on a part of the tree.

\begin{figure}[t!]
      \centering
      \includegraphics[scale=0.26]{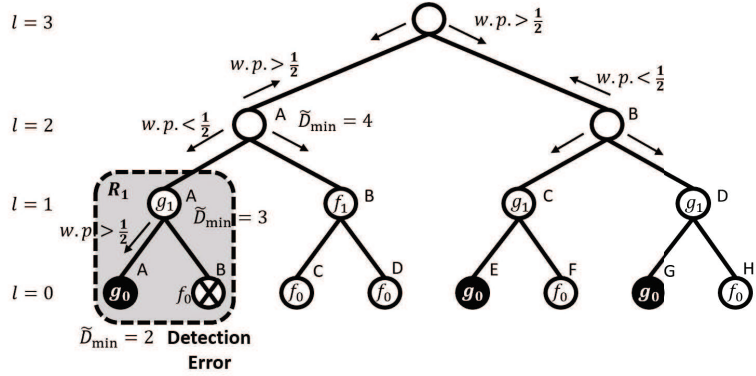}
      \caption{A biased random walk on the tree with detection errors.}
      \label{fig:error_tree}
   \end{figure}

In Fig.~\ref{fig:error_tree}, we illustrate an example with $M=8$, $L=3$, and $E=1$. Assume that after the first round of the test, there is a detection error that happened on level $l=0$, node $B$. In the next round of the test, when applying the IRW policy and starting from the root node, the probability of approaching the two targets on the right half tree is always greater than $\frac{1}{2}$. However, on the left half tree, due to the detection error, the observation on the higher level nodes would make the decision maker think that there are no more undeclared targets on the left half tree. The probability of approaching the left-most target is less than $\frac{1}{2}$ before the random walk enters the subtree $R_1$ as shown in Fig.~\ref{fig:error_tree}. But once the random walk enters the subtree $R_1$, the probability of approaching the left-most target becomes greater than $\frac{1}{2}$ since the detection error will not affect the observation from the true target anymore. In other words, for all the nodes below node $A$ on level $l=1$, the random walk will have a higher probability of approaching the left most target; for all the nodes above node $A$ on level $l-1$, the random walk will have a higher probability of leaving the left target. We call~$R_1$ the {\emph{affected subtree}}, the node $A$ on level $l=1$ the {\emph{changing point}}, and the left-most target the {\emph{affected target}}. 

For the general case, we provide the definition of these terminologies as follows. Since the affected trees may be in a nested structure,  they are defined in a recursive way. 

\begin{figure}[t!]
      \centering
      \includegraphics[scale=0.25]{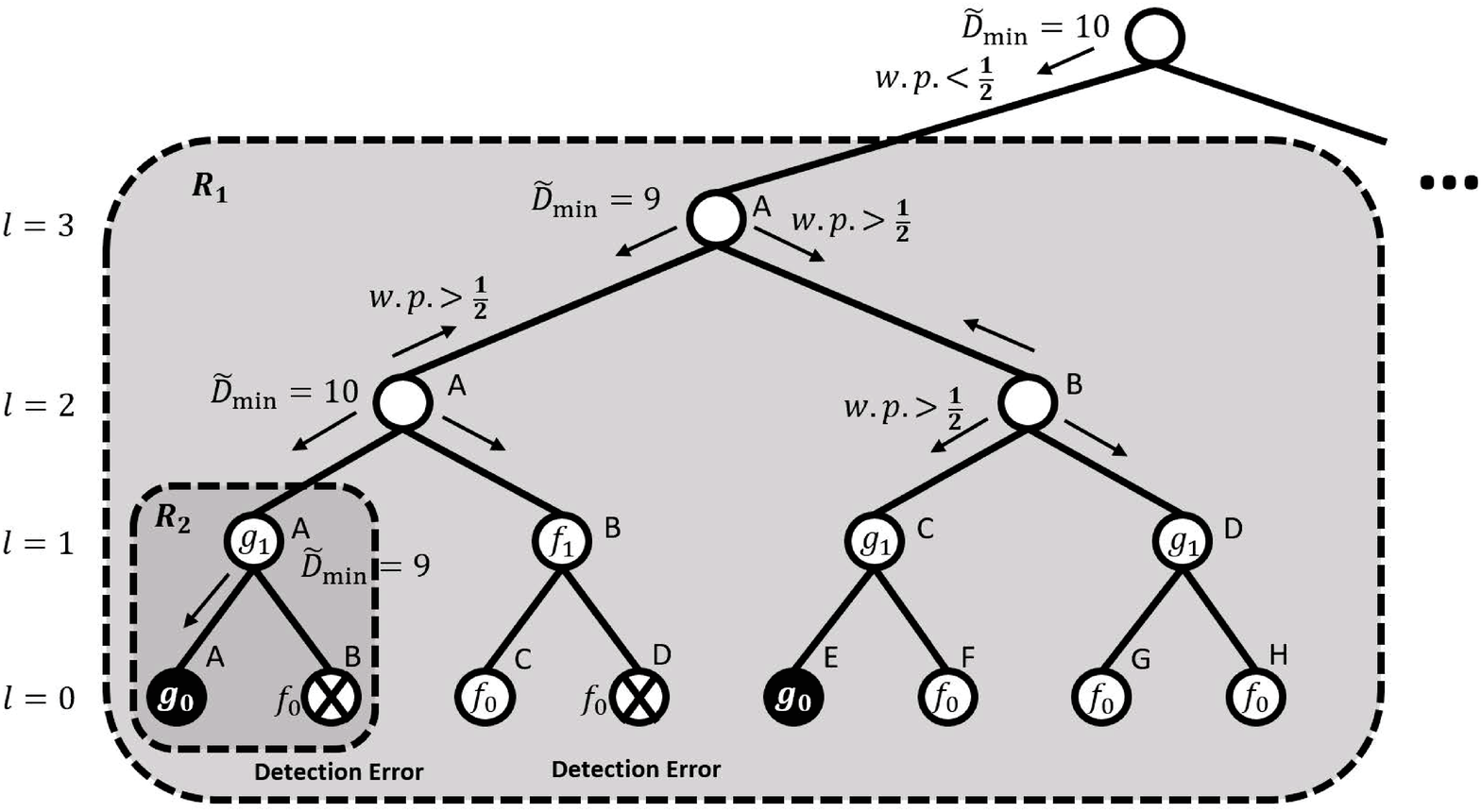}
      \caption{A biased random walk on the tree with detection errors: nested affected trees.}
      \label{fig:error_nested}
   \end{figure}

\begin{definition}
For a given tree structure, we define the affected subtrees from the lowest level to the highest level. 
We first define the affected subtrees rooted at level $l=1$. If an undeclared target has a detection error leaf as a sibling, the subtree formed by these two nodes and their parent node on level $l=1$ is defined as an affected subtree (e.g., $R_1$ in Fig.~\ref{fig:error_tree} and $R_2$ in Fig.~\ref{fig:error_nested}). By induction, after finding all the affected trees rooted at level $l=k$, a subtree rooted at level $l=k+1$ that satisfies the following two conditions simultaneously is defined as an affected subtree:\par
(1). There is at least one undeclared target node in the subtree which is not covered by any other lower level affected trees.\par
(2). The number of all the detection errors on the subtree is greater or equal to the number of all the undeclared targets on the subtree.   
\end{definition}

\begin{definition}
The roots of the affected subtrees are called {changing points}.
\end{definition} 

\begin{definition}
All the undeclared targets in an affected subtree are called {affected targets}. 
\end{definition}

There may be more than one affected subtrees in the detection and they are possibly in a nested structure. We illustrate another example in Fig.~\ref{fig:error_nested}, where $R_1$ and $R_2$ are two affected trees in a nested structure.

Our objective is to show that the detection delay of the IRW policy is upper bounded when there are detection errors in the tree. The proof idea is similar as before. We need to find a proper random variable that has a negative expectation (to approach the targets) at each step of the random walk. 

Let $\cV$ denote the set of all the target nodes; $\cC$ denote the set of all targets that have already been correctly declared; $\cA$ denote the set of the undeclared targets which are affected by the declaration errors; $\cU$ denote the set of the undeclared targets which are not affected by the declaration errors. It is easy to see that $\cC$, $\cA$ and $\cU$ are disjoint and $\cV = \cC \cup \cA\cup\cU$. 

For any node $v$ on the tree, depending on whether the node is on an affected tree, we consider the following two cases. 

Consider first that $v$ is not located in any affected trees. Define
\begin{equation}\label{error_dist_no_aff}
\tilde{D}_{\min}(v) := \min_{i\in \cU} \left\lbrace D_i(v)\right\rbrace,
\end{equation}
which is the minimum distance on the tree from $v$ to the undeclared targets which are not affected by the declaration errors.

Now consider that $v$ is located in an affected tree.
Since the affected trees may be in a nested structure, $\tilde{D}_{\min}(v)$ can be defined in a recursive way. Let $v_c$ denote the changing point of the affected subtree and $D_c$ denote the minimum distance from the change point to the undeclared targets on this affected tree. 

We define $\tilde{D}_{\min}(v)$ for the node $v$ from larger affected subtrees to smaller affected subtrees, from higher level to lower level. For the highest level changing point $v$ of the largest affected subtree, the parent of $v$ must not be on any affected trees, of which the $\tilde{D}_{\min}$ is defined in the previous bullet. We define a constant $Z$ as
\begin{equation}\label{add_on_Z}
Z := \tilde{D}_{\min}(\text{parent node of } v_c) - D_c -1.
\end{equation}
It is not difficult to see that $Z\geq 0$.

Within all the nodes on the current affected subtree which are not covered by any lower level nested subtrees, let $\cV_R$ and $\cV_T$ denote the sets of all tree nodes and all the undeclared targets,  respectively. For any node $v\in \cV_R$, $\tilde{D}_{\min}(v)$ is defined as
\begin{equation}\label{tilde_Dmin}
\tilde{D}_{\min}(v) = Z + \min_{i\in \cV_T} \lbrace D_i \rbrace. 
\end{equation}

For the nodes on all the lower level/nested affected trees, we use~\eqref{add_on_Z} and~\eqref{tilde_Dmin} recursively to find~$\tilde{D}_{\min}(v)$. It is not difficult to see that if there are no detection errors on the tree, $\tilde{D}_{\min}(v)$ coincides with $D_{\min}(v)$ defined in~\eqref{multi_dist}.

We now apply the definitions in~\eqref{add_on_Z} and~\eqref{tilde_Dmin} to provide examples to illustrate the proof. As shown in Fig~\ref{fig:error_tree}, $\tilde{D}_{\min}(v)$ of node $A$ on level $l=2$ is $4$ based on~\eqref{error_dist_no_aff}. For the affected subtree $R_1$, $Z$ equals $2$. Therefore, $\tilde{D}_{\min}(v)$ of node $A$ on level $l=1$ is $3$. For the example in Fig~\ref{fig:error_nested}, there are two affected subtrees $R_1$ and $R_2$. For $R_1$, $Z$ equals $6$. Therefore, $\tilde{D}_{\min}(v)$ for the node $A$ on level $l=3$ is $9$ and for the node $A$ on $l=2$ is $10$. For $R_2$, $Z$ equals $8$, which makes $\tilde{D}_{\min}(v)$ for the node $A$ on level $l=1$ be~$9$.

It is not difficult to see that after each step of the random walk, the variable $W_n =\Delta D_{\min}(v)$ will have the distribution
\begin{equation}
\begin{split}
\Pr \left(W_n\right) = \Pr \left(\Delta \tilde{D}_{\min}(v)\right)= 
\begin{cases}
p_l(v),  &\text{ for } W_n = \Delta \tilde{D}_{\min}(v)=-1,\cr
1-p_l(v), &\text{ for } W_n = \Delta \tilde{D}_{\min}(v)=1.
\end{cases}
\end{split}
\end{equation}
The IRW policy guarantees that $p_l(v)$ is always greater than $\frac{1}{2}$. Therefore, we have
$$\mathbb{E}[\Delta \tilde{D}_{\min}(v)] = 1- 2p_l(v) <0.$$

Similar to the sample complexity without detection errors, let $\tau_i$ denote the last time that the random walk has a distance greater or equal to $i+\log\frac{\log_2 M}{c}$ to all the targets. Therefore, after $\tau_i$ has elapsed, the random walk would have a distance less than $i+\log\frac{\log_2 M}{c}$ to at least one of the targets. However, by definition, the maximum value of $\tilde{D}_{\min}$ can be at most $2\log_2 M$. Using the same arguments as in the proof under the one-target case, we have that for all $\tau_i$ with $i = 1,\ldots, 2\log_2 M$, there exists a constant $\beta > 0$, such that 
\begin{equation}
\mathbb{E}[\tau_i]\leq \beta.
\end{equation}

Due to the constant $Z$ in the definition of $\tilde{D}_{\min}$ in~\eqref{tilde_Dmin}, the first state of the random walk might stop before $\sum_{i=1}^{2\log_2 M}t_i$. In this case, the detection delay of the random walk on the first state will still be bounded. Therefore, when there are detection errors on the tree, the detection delay $\mathbb{E}[\tau]$ of finding a target is upper bounded by 
$$\mathbb{E}[\tau] \leq 2B\log_2 M + \frac{\log\frac{\log_2 M}{c}}{D(g_0\|f_0)} + O(1).$$

With probability at most $O(Lc)$, the detection delay of finding all the $L$ targets with detection errors is upper bounded by:
\begin{equation}\label{multi_bound_error}
\mathbb{E}[\tilde{\tau}_{\text{all}}] \leq 2LB\log_2 M + \frac{L\log \frac{\log_2 M}{c}}{D(g_0\|f_0)} + O(L).
\end{equation}

By combining~\eqref{multi_bound} and~\eqref{multi_bound_error}, the detection delay is upper bounded by:
\begin{equation}\label{L_bound}
\begin{split}
\mathbb{E}_{\Gamma_{\mbox{\footnotesize IRW}}}\left[\tau\right]\leq (1-L\beta c)\mathbb{E}[{\tau}_{\text{all}}] +L\beta c \mathbb{E}[\tilde{\tau}_{\text{all}}] \leq LB\log_2 M + \frac{L\log \frac{\log_2 M}{c}}{D(g_0\|f_0)} + L^2B\beta c\log_2 M.
\end{split}
\end{equation}

In each round of the tests, the probability of detection error is upper bound by $O(c)$. By applying the union bound, the overall probability of error is bounded by 
\begin{equation}\label{multi_errors}
P_e \leq L\beta c = O(Lc).
\end{equation}
Combining~\eqref{L_bound} and~\eqref{multi_errors} completes the proof.

\bibliographystyle{ieeetr}
\bibliography{WangCohenZhao18TIT}

\end{document}